\documentclass[a4paper,12pt]{article}

\usepackage{amssymb,amsmath,mathbbol,mathrsfs}
\usepackage[usenames,dvipsnames]{color}
\usepackage{hyperref}
\usepackage{xcolor}
\usepackage{stmaryrd}
\usepackage{authblk}
\usepackage{framed}
\usepackage{empheq}
\usepackage{slashed}
\usepackage{hyphenat}
\usepackage{cite}
\usepackage{bm}
\usepackage{accents}
\usepackage{apacite}
\usepackage{natbib}

\usepackage{marginnote}    

\usepackage[left=.8in,right=.8in,top=.8in,bottom=.8in]{geometry}                  

\usepackage{sectsty}
\allsectionsfont{\sffamily\mdseries\upshape} 
\usepackage{tocloft}



\numberwithin{equation}{section}
5

\hypersetup{
	colorlinks=true,         
	linkcolor=MidnightBlue,          
	citecolor=BrickRed,        
	urlcolor=MidnightBlue            
}

\newcommand{\be}{\begin{equation}}
\newcommand{\ee}{\end{equation}}

\renewcommand{\d}{{\mathrm{d}}}

\newcommand{\pp}{{\partial}}

\newtheorem{theo}{Theorem}
\newtheorem{cor}{Corollary}

\newcommand{\cint}{{\int\kern-.87em{<}}}
\newcommand{\sint}{{\int\kern-.75em{\sim}}}
\newcommand{\fint}{{\int\kern-1.00em{\int}}}

\newcommand{\bb}{\mathbb}

\renewcommand{\#}{\sharp}

\let\oldmarginpar\marginpar
\renewcommand\marginpar[1]{\oldmarginpar{\color{red}\raggedright\footnotesize #1}}

\title{Geometrodynamics as Functionalism about Time}
\author{Henrique Gomes\footnote{\href{mailto:gomes.ha@gmail.com}{gomes.ha@gmail.com}} ~and Jeremy Butterfield\footnote{\href{jb56@hermes.cam.ac.uk}{jb56@hermes.cam.ac.uk}} \\\it Trinity College, University of Cambridge\\ \it Cambridge, CB2 1TQ, UK}

\begin{document}
\maketitle

We review three broadly geometrodynamical---and in part, Machian or relational---projects, from the perspective of spacetime functionalism. We show how all three are examples of functionalist reduction of the type that was advocated by D. Lewis, and nowadays goes by the label `the Canberra Plan’. 

The projects are: (1) the recovery of geometrodynamics by Hojman et al. (1976); (2) the programme of Schuller and collaborators (Schuller 2011; D\"ull, Schuller et al. 2018) to deduce a metric from the physics of matter fields; (3) the deduction of the ADM Hamiltonian by Gomes and Shyam (2016). 

We end by drawing a positive corollary about shape dynamics: namely, it has a good rationale for the Hamiltonian it postulates. \\ \\

\begin{center}
{\em Submitted to {\em From Quantum to Classical: Essays in memory of Dieter Zeh}, edited by Claus Kiefer: Springer, Cham, 2021.}
\end{center}

\newpage

  \tableofcontents

\newpage

\section{Introduction}\label{introFT2}
It is an honour to commemorate Hans-Dieter Zeh, whose contributions to foundational physics, especially in the areas of quantum decoherence and the nature of time, were so significant. We would like to do so by reviewing three broadly geometrodynamical---and in part Machian or `relational’---projects. The review is prompted by a recent literature in the philosophy of physics, about a doctrine called `spacetime functionalism’: which, as we will explain, each of the three projects vividly illustrates. Although these projects, and so this paper, are confined to classical physics, the final Section will briefly discuss quantum aspects. Thus we will briefly connect with Zeh’s own ideas about how his two main areas were linked: for example, that the matter degrees of freedom could decohere the metric-gravitational degrees of freedom so as to enable, in an Everett-style interpretation, a classical time to emerge. 

We first introduce the projects (Section \ref{3projects}). Then we give some details about the physics  (Section \ref{appetizer}), and the philosophy (Section \ref{funcredn}). These details come together in Section \ref{3illustr}.

\subsection{Three projects}\label{3projects}
Recently, a literature has sprung up about `spacetime functionalism’. Like all `isms’, it comes in various versions. But the broad idea is that the concept of spacetime, and other more specific chrono-geometric concepts such as `distance’ `simultaneity’ or `inertial frame’, are functional concepts. This means, roughly speaking, that the concept is defined by its pattern of relations to other concepts. The pattern is called the concept’s {\em functional role}; so the idea is that you best understand the concept by looking at `the role it plays’. Hence the slogan: `spacetime is as spacetime does’. Here the `does’—what makes up the functional role---is realized i.e. instantiated by the physics of matter and radiation. So spacetime functionalism is closely related to relational, and specifically Machian, approaches to chrono-geometry and dynamics; and to what has recently been called the ‘dynamical approach’ to chrono-geometry.\footnote{Spacetime functionalism has also recently been a theme in philosophical discussions of the emergence of spacetime from quantum gravity. We will here set this aside, apart from our brief return to Zeh in our final Section. But we thank Nick Huggett for stressing to us the question whether the concepts in a putative quantum gravity theory---the ``bottom’’ theory in Section \ref{funcredn}’s jargon---can be grasped without invoking spatiotemporal notions. For more discussion, cf. e.g. \citet[Introduction, Section 6]{HuggettWuttrich_book} and \citet{LamWuttrich_func}.}

We are sympathetic to spacetime functionalism. But in two previous papers, we argued that the recent literature missed a trick---though there was also good news. 

The missed trick was that this literature did not notice that in its original and, we think, best formulation (by Lewis), functionalism is a variety of reduction. In {\em reduction}, a problematic discourse or theory is vindicated or legitimized by being reduced to (i.e. shown to be a part of) an unproblematic discourse or theory. In {\em functionalism}, this reduction is achieved by spelling out the functional roles of the problematic discourse's concepts or properties, and arguing that each of them is uniquely realized i.e. instantiated  by certain (usually complicated) concepts or properties of the unproblematic discourse.  

Thus we urged that spacetime functionalism should say that a chrono-geometric concept is, or several such concepts are, uniquely specifiable in terms of their functional roles (patterns of relations). Then the thrust of the reduction---and the echo of Machian and relationist ideas---is that these functional roles are realized by (usually complicated) features of the physics of matter and radiation, without adverting to other concepts of chrono-geometry. 

The good news was that the older literature in foundations of chrono-geometry (before the recent label ‘functionalism’) contained successful examples of this reductive endeavour. We reported four such examples: cases where the unique specifiability of a chrono-geometric concept in terms of matter and radiation, and the corresponding reduction, was secured by a precise theorem.\footnote{\label{OurPrevious}{The `missed trick’ accusation, and our overall account of functionalism and reduction, is in our \citet{ButterfieldGomes_1}. The `good news’ examples are in our \citet{ButterfieldGomes_2}. Of these examples, the fourth is Barbour and Bertotti’s Machian dynamics \citet{Barbour_Bertotti}, in which a temporal metric is defined by the dynamics of the point-particles. So this example also illustrates this paper’s specific topic, of functionalism about the temporal metric.}}

We can now state this paper's aim: to show that three projects in the physics literature give vivid and impressive illustrations of spacetime functionalism. As it happens, they have hitherto been almost wholly ignored by philosophers of physics;  so we submit that they deserve philosophers' attention. The first project is forty years old and is well-known in mathematical relativity. But the other two are much more recent and less well-known---so, again: worth advertising.  

More specifically, each of these projects illustrates functionalism {\em about time}. So in broad philosophical terms, they each assume that:\\
\indent \indent (a) some discourse or theory about matter and radiation, and  {\em space}, is unproblematic; while \\
\indent \indent (b) a discourse or theory about time is problematic, and is to be vindicated by being reduced to (shown to follow from) the unproblematic.\\
In slightly more technical terms:  each provides, within a theory about spatial geometry, a functionalist reduction of the temporal metric and time-evolution: and the reduction is summed up in a theorem that the temporal metric and-or the Hamiltonian governing time-evolution is, in an appropriate sense, unique.\footnote{The words `problematic’ and `unproblematic’ are our labels: these projects’ authors do not use them. We admit they are imperfect: one might prefer less judgmental adjectives such as `troublesome’ and `okay’ (a suggestion we discuss in Section \ref{funcredn}). And whatever the adjectives, one could of course resist the urge to philosophize! One could abjure all ideas about reducing chrono-geometry, and read the uniqueness results as ``just’’ mathematical theorems. But as we read these projects’ authors, they {\em do} sympathize with the endeavour of reduction.}

These three lines of work are all `general-relativistic', in a sense we will make precise. But they differ substantially in exactly what they assume, and in what they deduce. They are, in short:\\
\indent \indent  (1): The recovery of geometrodynamics, i.e. general relativity's usual Hamiltonian, from requirements on deformations of hypersurfaces in a Lorentzian spacetime. This is due to \citet{HKT}.\\
\indent \indent  (2): The programme of Schuller et al.: \citet{Schuller, Dull}. They deduce from assumptions about matter and radiation in a 4-dimensional manifold that is {\em not} initially assumed to have a Lorentzian metric, the existence of a `generalized metric', with e.g. causal future and past cones and mass hyperbolas; and they deduce a lot about how their generalized metric relates to matter and radiation. \\
\indent \indent  (3): The deduction of general relativity's usual Hamiltonian in a framework without even a spacetime: that is, without initially assuming a 4-dimensional manifold, let alone one with a Lorentzian metric. This is due to \citet{GomesShyam} (with  precursors in  \citet{Giulini1995} and \citet{Barbour_RWR}).

We will discuss these in this order: in Sections \ref{sec:HKT}, \ref{sec:DGS} and \ref{sec:GS} respectively. Then in Section \ref{sec:payoff}, we give a positive corollary of (3). In short: (3)  answers a misgiving you might have about a recent programme in the foundations of classical  gravity, viz. shape dynamics. Section \ref{sec:concl} concludes with a table summarising the results surveyed, and briefly draws some connections with the work of H-D. Zeh. 

\subsection{Appetizer: the projects introduced}\label{appetizer}
 
All three projects are relativistic, indeed general-relativistic. In calling them `general-relativistic', we mean, more precisely: (a) they either assume or deduce a Lorentzian metric, and a dynamical geometry, i.e. a spatial geometry that varies in time according to an equation of motion; and (b) they focus on this equation of motion being the orthodox general-relativistic one.  Thus in this paper, the label `general relativistic' connotes a notion of evolution, and, indirectly, of global hyperbolicity, that is not present in the Einstein field equations {\em simpliciter}.

But note our phrases: `either assume or deduce'; and `focus on', rather than `prove'. That is, as these phrases suggest:  these projects' details differ considerably. There are contrasts both about what is assumed, and about what is proved.  For example, each of the first and third projects deduces the orthodox general-relativistic Hamiltonian in the presence of any of a wide class of matter fields; but the second project does not. And while the first project assumes a 4-dimensional spacetime manifold and Lorentzian metric, and the second project assumes a spacetime manifold: the third manages to avoid assuming either of these concepts. It assumes only a spatial 3-manifold, from which it in effects {\em builds} a spacetime manifold. 

We introduce the three projects, in more detail as follows.

\begin{enumerate}
\item The first project \citet{HKT} assumes a 4-dimensional spacetime manifold with a Lorentzian metric. But Hojman et al. adopt a `3+1' or `canonical' perspective: so they consider 3-dimensional spacelike hypersurfaces each equipped with its Riemannian geometry, and discuss how such a space, $\Sigma$ say, can be embedded in spacetime. They show that if we transcribe some natural requirements about how these embeddings mesh---the requirements are encoded in a {\em  deformation algebra}---into a Hamiltonian framework for describing how the spatial geometry on $\Sigma $ changes, so as to give a constraint algebra, then the Hamiltonian {\em must be} the usual Hamiltonian in general relativity in its ADM form. This is a remarkable, and rightly lauded, achievement.

In short, and in philosophers' jargon: Hojman et al. show that the ADM Hamiltonian is the unique realizer of the functional role `... is the generator of how spatial geometry changes over time, that reproduces the assumed deformation algebra of hypersurfaces'. 

The technical apparatus used by this first project is also important for us, since parts of it will appear again in the second and third projects.

\item On the other hand, the second project \citet{Schuller, Dull} assumes a spacetime manifold that is {\em not} considered as foliated, nor is even required to have a Lorentzian metric. It then proceeds in two stages. The first stage assumes that the dynamics of any matter-or-radiation fields in the spacetime obey some conceptual requirements, e.g. about being `well-behaved' and `predictive' in certain senses, and about being deduced from a diffeomorphism-invariant action functional. These assumptions imply a  description of a causal structure for the spacetime, and so of spatial hypersurfaces, and of how a generalized, tensorial spatial geometry changes over time, satisfying certain properties. Then the second stage adopts a 3+1 decomposition, and applies the apparatus of Hojman et al., i.e. the first project,  to show that this description is consistent, i.e. can be satisfied. 

So far, the only specific matter-or-radiation dynamics for which the resulting consistency equations have been solved is classical electromagnetism: for which the resulting evolution of spatial geometry is (like in the first project) exactly as described by general relativity, i.e. by the ADM Hamiltonian. (In saying `the only known solution', we do not a criticism of the programme: it is a familiar point that a reduction can be of restricted scope, a `local reduction’, yet of great scientific interest.)

This is, we submit, a remarkable achievement. It deserves to be better known by philosophers: not least because of its evident affinities with the `dynamical relativity' viewpoint of \citet{Brown_book}, and the spacetime functionalism literature (which was prompted in part by that viewpoint). Thus in this project, the functional role of the temporal metric requires that the equations of motion of the matter-or-radiation fields should be well-behaved and predictive in senses that turn out to define causal cones and make time ``the dimension along which we make predictions.''\footnote{This project's requirement that the dynamics of matter be predictive yields another affinity. Namely, with the proposal \citet[Ch. 7]{Callender_book}  that time is characterised in contrast to space by the fact that data can be specified on spacelike surfaces, but not timelike and null surfaces, in an unconstrained way. He argues that this makes partial differential equations that evolve data from spacelike surfaces `more informative', in a sense that he derives from a broadly Humean account of laws of nature as  summaries of informative patterns in phenomena. We will not pursue details of this imaginative proposal: \citet{James2020} is a detailed assessment.} In Section \ref{sec:DGS}, we will discuss these affinities in more detail. 

For the moment, we just note the obvious moral: that here is another uniqueness result {\em \`{a} la} functionalism. In short, it is: given these authors' general assumptions, and classical electromagnetism, general relativity i.e. the ADM Hamiltonian uniquely satisfies the functional role `... describes how spatial geometry changes over time' (where this last phrase gets spelt out in terms of predictivity etc.). 

But there is a {\em caveat}: these authors' deduction of general relativity depends sensitively on their assumption of spacetime diffeomorphism invariance. For we will see (in Section \ref{sec:counter}) that without it, there are models of their other assumptions that have the Maxwell field coupled to the Lorentzian metric in the usual way, and obeying Maxwell's equations (in temporal gauge), while the evolution of geometry is completely different from that of general relativity. More generally: it is not clear how broad a class of assumptions within these authors' framework, would yield orthodox couplings between matter or radiation, and a Lorentzian metric with orthodox i.e. general-relativistic evolution of geometry.

\item Finally, the third project \citet{GomesShyam} is more austere, i.e. self-denying, in its initial assumptions about the `unproblematic' base, from which to build the reduction of the `problematic'  temporal metric. It does {\em not} assume, {\em ab initio}, a spacetime manifold. It assumes a 3-manifold whose spatial geometry is to vary over time, in a manner that is locally definable in a certain sense: roughly, that local changes of a quantity in distinct spatial regions are required to commute in an appropriate way. This assumption is then set in a Hamiltonian framework, with some technical conditions, such as the Hamiltonian being at most second-order in momenta.

Remarkably, it turns out that these assumptions are enough to {\em construct} a spacetime, with a  Lorentzian metric; {\em and} to show that the geometry evolves over time according to the general-relativistic Hamiltonian, i.e. the ADM Hamiltonian. Besides, this construction holds good in the presence of any of a wide class of matter and radiation fields, not just e.g. electromagnetism.\footnote{This project, as well as project (1) of Hojman et al., naturally allow for an Euclidean version of geometrodynamics. In both cases, we can impose further conditions that would specialize to the Lorentzian signature. }

We will see in Section \ref{sec:GS} that this result is a descendant, for relativistic field theories with a dynamical spatial geometry, of Barbour and Bertotti's definition of the temporal metric in their  non-relativistic Machian particle dynamics (\citet{Barbour_Bertotti},  reviewed in our \citet{ButterfieldGomes_2}; cf. footnote \ref{OurPrevious}). 
 
So here again is a uniqueness result, consonant with functionalism. The ADM Hamiltonian is the unique realizer of the functional role `... is the generator  of how spatial geometry changes over time, in a locally definable way, without assuming spacetime'. Broadly speaking, this is a stronger result than predecessors since there is no initial assumption of a spacetime.

Besides, we will see in Section \ref{sec:payoff} that this result has a positive corollary for a recent programme in classical  gravity, viz.  shape dynamics. Namely: this result exonerates shape dynamics from the accusation that, because in a certain regime its Hamiltonian matches that of general relativity, shape dynamics  amounts to  `theft' rather than `honest toil', as Bertrand Russell's much-cited quip puts it. In a bit more detail, the point will be: agreed, shape dynamics can admit to having, as a matter of its own history, `piggy-backed' on general relativity in the formulation of its Hamiltonian. But this result shows that this `piggy-backing' need not be `theft'. For shape dynamics can `take over'---i.e. an advocate of shape dynamics can transcribe into their formalism---the derivation of Section \ref{sec:GS}'s result, and so obtain the desired Hamiltonian---even without initially assuming a spacetime manifold.  In short, the result has the merits of `honest toil', and shape dynamics can invoke that toil, to answer the accusation of `theft'.\\

 \end{enumerate}

To sum up this appetizer: all three projects have as their punchline, an impressive uniqueness result, that the orthodox Hamiltonian of 3+1 general relativity, the ADM Hamiltonian, is the unique satisfier of a condition. In philosophers' jargon: it is the unique realizer of a functional role.  The results vary in their assumptions. In particular: as one passes from (1) to (3), less is assumed about the manifold. One ends with (3) only assuming a 3-manifold with a spatial metric; and while (2) assumes a 4-manifold, it does not postulate {\em ab initio} any metric, even a spatial one. Also, the detail of what is deduced varies. For example, both (1) and (3) apply to a wide class of matter fields; while for (2) the uniqueness result is, so far, `local', i.e. proven only for electromagnetism.

\subsection{Functionalist reduction: a review}\label{funcredn}
In Section \ref{3projects} we said about reduction and functionalism, only that: \\
\indent \indent (i) in a functionalist reduction, the functional roles of the concepts or properties of the problematic discourse or theory are argued to be uniquely realized by concepts and properties of the unproblematic discourse or theory; and \\
\indent \indent (ii) for this paper’s three projects, it is time, especially the temporal metric, that is problematic, while spatial geometry, and the physics of matter and radiation, are unproblematic. \\
In this Section, we will fill out (i). We will begin with how philosophers of science (and indeed, philosophically inclined physicists) usually discuss reduction of one theory to another. This will be a matter of recalling the basic idea of Nagel’s account of reduction. Then we will expound functionalism, in the formulation of Lewis. We will introduce functional definition, and then stress how functionalist reduction differs from---we say: improves on---Nagelian reduction. (All this will summarise Sections 2 to 5 of our \citet{ButterfieldGomes_1}: they contain various developments, defences and references which, for reasons of space, we here forego.) With this review in hand,  Section \ref{3illustr} will then spell out how this paper’s three projects illustrate functionalist reduction, thus filling out (ii) above.

\paragraph{Nagelian and functionalist  reduction contrasted}~
In {\em Nagelian  reduction}, one envisages reducing one theory,  say $T_t$ (`t' for `top'),  to another,  say $T_b$ (`b' for `bottom'), by adding to $T_b$ a set $B$ of so-called {\em bridge-laws}. The core idea is that these are sentences that use the vocabularies of both $T_b$ and $T_t$ in such a way that from the conjunction of $T_b$ and $B$, one can deduce all of $T_t$ (\citet[pp. 354-358]{Nagel_book};  \citet[pp. 361-373]{Nagel_1979}). So, assuming that theories are sets of sentences closed under deduction, $T_t$ is shown to be part of, i.e. already contained in, $T_b$. The traditional idea is that $T_b$ is an improvement on $T_t$, since it recovers $T_t$’s successes but also says more. So `b’ is also a mnemonic for  `better’ and `t’ is a mnemonic for  `tainted’.\footnote{\label{conversejargon}{Beware: philosophers and physicists use `reduce’ in converse senses. We adopt the philosophical jargon, in which the `tainted’ or `worse’ theory is reduced to the better one; but physicists say that, for example, special relativistic kinematics reduces to Newtonian kinematics for small velocities.}} 
   
Several adjustments or variations of this core idea are widely agreed (including by Nagel himself: \citet{Frigg_Nagelian, Butterfield_emergence, SchaffnerNagel2012}).  Most philosophers  agree that to fit real-life scientific cases, one must allow that, not all of $T_t$, but only most of it, or some approximation or analogue, of it, is deduced from the conjunction of $T_b$ and $B$. And most philosophers agree that bridge-laws do not always (even collectively) `define' each $T_t$ term, even in the usual minimal sense of  logic books: viz. specifying the term's (contingent) extension---nevermind its intension or ``meaning''. (For a predicate $F$, the extension is the set of actual instances of $F$; for a name or other expression referring to an object e.g. a definite description `the unique $F$’, it is that object.) After all: not determining such extensions is perfectly compatible with our nevertheless being able to deduce all of $T_t$---as Nagelian reduction demands. 

This ``looseness'' means that there can be a choice of which bridge-laws to postulate: a choice of which set $B$ to add to $T_b$, in order to deduce $T_t$, or an approximation or analogue of it. For there might be more than one way to relate all of $T_t$'s terms to those of $T_b$, while nevertheless securing the deduction. Of course, this is not to say that all such choices are equally good; and one would expect it to be a matter of scientific judgment, relative to perhaps various criteria, which bridge-laws to postulate. Indeed, most philosophers in the Nagelian tradition regard it as a matter of scientific judgment, whether  to postulate  bridge-laws at all and thereby perform a reduction. Thus their rationale for a reduction is often that it makes for a more unified, or more parsimonious or simpler, picture of the world.

We can now state the contrast with {\em functionalist reduction}, as formulated by Lewis. It says instead:\\
  \indent \indent (i): Each bridge-law connects just one $T_t$ term with the vocabulary of $T_b$.\\
 \indent \indent (ii):  Each bridge-law `defines' its $T_t$ term,  in logicians' weak sense of specifying its extension. But beware: (a) the extension specified by the bridge-law is contingent,  and need not reflect pre-given meanings of the terms involved; and (b) the bridge-law  `definition' is {\em not} one of the functional definitions emphasised by functionalism---for as we will see in a moment, the functional definition of a $T_t$ term is given {\em wholly within} $T_t$.  \\
 \indent \indent (iii):  The bridge-laws are contingent statements: they are statements of contingent co-extension. (That is: two predicates have the same set of actual instances; two referring expressions actually refer to the same object.) But they should {\em not} be called `hypotheses'. For they are {\em mandatory}, not optional. There is no variety or choice or optionality, of the sort Nagelians envisaged.

Lewis defends these proposals about reduction and bridge-laws as part of his overall view about {\em functionalism}. But in his classic expositions of his overall view \citet{Lewis_defT, Lewis_func}, these proposals about bridge-laws come after his main statements about functionalism, i.e. towards the ends of the papers (viz. at \citet[pp. 441-445]{Lewis_defT}; \citet[255f]{Lewis_func}). As a result, in the subsequent literature they have been much less emphasised than the main statements. This has been unfortunate since it has fostered the widespread view that functionalism is opposed to reduction---when in fact Lewis has shown how they can be persuasively combined.\footnote{\label{Canberra}{The worthy exception to this unfortunate neglect is `the Canberra Plan’. This is the nickname for advocacy of Lewisian functionalist reduction, which is familiar in metaphysics, philosophy of mind and ethics (to which Lewis also applied the doctrine): but which is unfortunately almost unknown to philosophers of science, and wholly unknown to recent `spacetime functionalists’. So our position in this and our other papers is that philosophy of chrono-geometry gives good illustrations of the Canberra Plan. As will be clear, our main difference from the usual Plan will be that we do not require the functional tole of a concept, extracted from the top theory $T_t$, to be strictly faithful to some pre-given meaning, or to give what philosophers call a `conceptual analysis'. For details of the Canberra Plan, cf. \citet{MitchellNola}.

For physicist readers, a further word of introduction about Lewis might be helpful. He was a giant of twentieth-century philosophy; and although he never worked in philosophy of physics, his views about topics other than reduction (especially about modality and causation), have close connections with foundational issues in spacetime theories, and in mechanics, both classical and quantum. Cf. for example, \citet{Butterfield_hole, Butterfield92, Butterfield_DLHJ}}.}   

To explain Lewis’ proposals (i) to (iii), the key idea is that while Nagelian reduction contrasts  two vocabularies, viz. those of the two theories $T_t$ and $T_b$: for Lewisian reduction, there are {\em three} vocabularies at issue---not two. The reason is that {\em within} $T_t$, there is a division of vocabulary, into two sorts. Each term in one sort will get {\em functionally defined} in terms of items in the other sort. So for the next few paragraphs, we set aside reduction, and focus only on $T_t$.

\paragraph{Functional definition}~
First, there are the terms\footnote{We concentrate on predicates. Let us assume that names and other expressions referring to objects, e.g. definite descriptions, are eliminated in favour of predicates in the usual way.} that have an agreed interpretation in a certain domain.  Lewis calls them $O$-terms, where $O$ stands for `old'.  Another good mnemonic is `okay' (suggested by \citet[p. 55]{ButtonWalsh}). For the idea is that $O$-terms are fully understood: there is nothing problematic about them. So as Lewis stresses: $O$ does not stand for `observational', as against `theoretical' in the sense of `non-observational'.\footnote{\label{LewNotInstruml}{So Lewis does not aim to `reduce theory to observation' in the way that some logical empiricists did. More generally: his treatment of both functionalism and reduction is ``realist'', not ``instrumentalist'' or ``eliminativist'' about $T_t$.}} So there is an agreed, unproblematic or `okay', notion of truth in that domain for sentences of $T_t$ that contain only $O$-terms. 

Then there are the other terms of $T_t$:  terms that, Lewis proposes, $T_t$ introduces to us for the first time, so that their interpretation is not settled.  Lewis calls them $T$-terms. But again, as he stresses: $T$ does not stand for `theoretical', as against `observational'---any more than his $O$ stands for  `observational'. The $T_t$-terms are just {\em new} terms, whose interpretation needs to be settled. Another good mnemonic is `troublesome' (Button and Walsh ibid.): for the $T_t$-terms are troublesome, problematic, at least in the minimal sense of our needing to interpret them.

Lewis now proposes that the assertions of $T_t$, taken together, give sufficiently rich information about how all the $T$-terms are related, both to the (old, interpreted) $O$-terms in $T_t$, and to each other, that the truth of $T_t$ (and so advocacy of $T_t$) implies that each $T$-term has a {\em unique} interpretation. That is: advocacy of $T_t$ involves claiming that each such term has a unique extension, that is implicitly determined by the entirety of $T_t$.  Assuming this uniqueness, Lewis then provides a systematic procedure whereby given $T_t$, one can write down definitions (in the usual logicians' weak sense of: specifications of extensions) of each of the $T$-terms. So he provides a procedure for getting {\em simultaneous unique definitions}. 

Here we meet the jargon of {\em functional role}, announced at the beginning of Section \ref{3projects}. The general idea is that a property or concept of interest can be uniquely specified by a pattern of relations (typically: causal or nomic, i.e. law-like, relations) that it has to other properties. The pattern is called the property's {\em functional role}. The property is called the {\em realizer} or {\em occupant} of the role.\footnote{(1): Nothing here will turn on any distinction one might make between property and concept. So we will mostly say `property'. But note that accordingly, `properties' includes also relations, of two or more places. (2):  We will mostly say `specified', `specification' etc., though often `defined' and `definition' is used. Saying `specified' and `specification' has the advantage of avoiding the connotations of either free verbal stipulation, or of being faithful to a pre-given meaning. This is an advantage since in some examples these connotations do not hold good.} 

In this paper, we do not need the details of Lewis’ procedure for extracting from $T_t$ specifications of the extensions of each of its $T$-terms: his procedure for getting {\em  simultaneous unique definitions}. The reason is essentially that within each of our three projects, only one $T$-term, viz. the temporal metric, gets specified. So for us it suffices to note the following six points.\\
\indent \indent (1): When we envisage a whole set of properties being functional, i.e. specified by their relations to each other and to yet other  properties, one naturally worries that there will be a vicious ``logical circle'' of specification. For if one property $X$ is specified by its functional role that mentions another property $Y$, {\em and} vice versa, i.e. $Y$'s functional role mentions $X$: then surely there is a circle, and both specifications fail. And similarly for circles with more than two members. \\
\indent \indent (2):  But Lewis showed that there need be no logical circle. To avoid such circles, one need only maintain that the body of doctrine mentioning the properties (in our notation: the theory $T_t$) is sufficiently informative or rich (logically strong) that each of the functional roles is satisfied by just one property. That is: one maintains that each functional role has a {\em unique} realizer. Then each such property can be specified by its functional role, without any logical circularity. Each $T$-term has its extension fixed by its functional role; and although the functional role seems to include other $T$-terms, they are eliminated, so that {\em au fond} the functional role uses only $O$-terms.\footnote{\label{cluedo}{For details, cf.:\citet[pp. 428-438]{Lewis_defT}, with ancillary material in 438-441; \citet[pp. 253-254]{Lewis_func}; our \citet[Section 4]{ButterfieldGomes_1}. Lewis also gives (1972, 250-253) as a parable, a detective story in which the detective gathers information about three conspirators to a murder, whom he labels $X, Y$ and $Z$: sufficiently rich information that for it to be true, $X, Y$ and $Z$  must be three people, whom he and the police know by other means---so that they can be arrested.}} \\
\indent \indent (3): The main area in which this insight, of simultaneous unique definability,  has been discussed is philosophy of mind. Here, mental properties are considered problematic; while bodily and behavioural properties are unproblematic.  So the functionalist idea is that each mental property, like being in pain or believing that it is raining, is specified by its functional role: its characteristic pattern of relations to other mental properties, and to bodily/behavioural properties. For example: being in pain is specified by its being typically caused by tissue damage and its typically causing both distress (another mental property) and aversive behaviour. So here, $T_t$ is the non-technical everyday theory of mental lives and behaviour, often called `folk psychology'. And in this area, Lewis' insight is that by extracting mental properties' functional roles from folk psychology, these properties can all be simultaneously uniquely specified, without logical circularity, in terms of bodily and behavioural properties.\\
\indent \indent (4): Lewis does not require unique realization in ``all possible worlds'', but only in the actual world: `I am not claiming that scientific theories are formulated in such a way that they could not possibly be multiply realized. I am claiming only that it is reasonable to hope that a good theory will not in fact be multiply realized' \citet[pp. 433]{Lewis_defT}.\\
\indent \indent (5): Lewis of course discusses the cases where there is not a unique realizer, or more than one; and also nuanced cases like (a') there is none, but there are near-realizers, of which perhaps one is nearer than the others; and (b') there are many, but some are (perhaps one is) in some way a better realizer than the others. (Cf. also \citet[Sections 3.3-3.5, 4.2 and 5]{ButterfieldGomes_1}.) But here we can skip details, since this paper’s three projects are `clean'. In each of them, the functional role of the temporal metric has a unique realizer. (But we will briefly meet a nearest realizer in Section \ref{sd_exon}.) \\
\indent \indent (6): Lewis is not suggesting that $T_t$ says all there is to know about the interpretation of a $T$-term. On the contrary: we might come to know much else about it, after or independently of our acceptance of $T_t$.  This point is all-important for Lewis’ account of reduction and bridge-laws---to which we now return.

\paragraph{Functionalist reduction and mandatory bridge-laws}~
Suppose that after accepting, or independently of accepting, $T_t$ and its simultaneous functional specifications, we come to accept {\em another} way of specifying the realizers of the functional roles. Then since the realizer of each functional role is unique, we {\em must} accept an identity statement---we must identify the original realizer with the new specification.  In the jargon of reduction: we must accept that the original property is reduced to the new specification. 

In terms of our notation of $T_t$ and $T_b$, and the example of pain (cf. point (3) above):--- Suppose we develop a theory $T_b$ of neurophysiology. It has its own specialist vocabulary: `neuron’, `synapse’ etc. But it may well also use some or all of the unproblematic $O$-terms (bodily and behavioural terms) of $T_t$, i.e. of folk psychology. Then for the example of pain: we can imagine discovering that a neurophysiological property, say `C-fibre firing', realizes the functional role of pain, as it was spelt out in the $T_t$. Then we {\em must} accept: pain is C-fibre firing. This is the point that  we labeled (iii), when we contrasted Nagelian and functionalist reduction. Namely: in the latter, bridge laws are mandatory. 
 
Now we also see why there are {\em three} vocabularies at issue in reduction; (not two, as discussions of reduction usually assume---one for $T_t$, one for $T_b$). For first: the reduced theory's vocabulary divides into two sorts, a division that sets the stage for functionally defining the first in terms of the second; and then there is the reducing theory's vocabulary---in our example, neurophysiological vocabulary (`neuron’, `C-fibre firing').
 
These ideas are all illustrated in this inference, from two premises: \\
\indent \indent \indent  i):	Being in pain = the inner state that is typically caused by tissue damage and typically causes both distress and aversive behaviour;\\
\indent \indent \indent ii)	C-fibre-firing = the inner state that is typically caused by tissue damage and typically causes both distress and aversive behaviour; .\\
\indent \indent \indent So, iii): Being in pain = C-fibre-firing. \\
We accept i) because of $T_t$. This first premise is the functional definition, with `distress'  being, like `pain', a mental term; whose occurrence in the {\em definiens} signals the idea (and legitimacy!) of simultaneous definition.  We accept  ii) by accepting the later (or independent) theory that we have labelled $T_b$: for this example, contingent neurophysiology. And we infer iii) by the transitivity of identity (i.e. of co-extensiveness of predicates). Thus iii) is the {\em derived bridge-law}. It is a contingent statement of co-extension (since premise i) is contingent).

To sum up: this example makes clear how we are, as Lewis puts, `logically compelled to make theoretical identifications' 1970, p. 441), given premises to which we are already committed. To put the premises in general terms: \\
\indent 1) according to $T_t$, our $T_t$-term $\tau$ (e.g. `being in pain')  has a unique realizer (extension, referent), that is specified by its pattern of relations, its functional definition; and \\
\indent 2) according to our later (or independent) theory $T_b$,  $\rho$ (`$\rho$' for `reducing') also refers to that realizer.\\
So we must infer: $\tau = \rho$.

It seems that many scientific examples of reduction fit this account. An obvious, because famous, example is Maxwell's theoretical identification of visible light with electromagnetic waves. To put this with a cartoon-like simplicity: recall that Maxwell calculated within his electromagnetic theory that some solutions were wave oscillations of the electric and magnetic fields, that travelled at ca. $3 \times 10^5$ km s$^{-1}$. This was very close to the measured speed of visible light; and it strained credulity to imagine that two mutually independent phenomena propagated at the same speed. So he inferred that light consisted of these oscillating solutions to his equations. In terms of the notation above:  the (rough!) functional role is `travels at ca. $3 \times 10^5$ km s$^{-1}$'; the $T$-term `visible light' in the contemporary theory of optics ($\tau$ in our $T_t$) is functionally defined by this role; and `(suitable) oscillating solution of the Maxwell equations' is the other term $\rho$ drawn from the reducing theory $T_b$, i.e. Maxwell's electromagnetic theory. 

Agreed, this is a {\em cartoon} of what Maxwell did.  And one naturally fears that many, perhaps most,  real-life scientific cases that are called `reductions'  will not conform to this neat account. But  there are two points to make in its defence; the second is specific to this paper.\\
\indent \indent (i):  Lewis also gives two examples fitting his account  (1970, 443-444),  argues that other cases do conform to it (1970, 444-445), and also discusses how the account treats later revisions of the theories (1970, 445-446).\\
\indent \indent (ii):  As we said in (5) above: Lewis also discusses what to say in cases where there is not a unique realization, or more than one. (Indeed, we now know that in the Maxwell example, there is more than one realizer of the functional role, `travels at ca. $3 \times 10^5$ km s$^{-1}$': gravitational influence, as well as light.) But in this paper, we can skip these complications, since our three projects give `clean' examples---of time having a functional definition, and a reduction.  \\

\subsection{How the projects illustrate functionalist reduction}\label{3illustr}
To prepare for these projects and to summarise this Section, we end by stating the trio of claims that will be the analogues of premises i) and ii) and the conclusion iii) above. Indeed, we will do this twice: first, in general, indeed vague, terms, so as to fit equally well all three projects; and then, in more specific terms that differ between the projects. 

Note that our giving both a general trio, and more specific trios, is not a matter of indecision. For our functionalist reductive perspective is a philosophical template that can be brought  into contact with the scientific details in various ways. And we think there need be no---there probably is no---fact of the matter about which among the various precise $T_t$ one could diligently extract from e.g. \citet{HKT} is the best, or the right, $T_t$ to fix on as providing a functional definition of the time parameter (a definition that is then realized by properties mentioned by a corresponding Hamiltonian `bottom theory’ $T_b$). 

So first, the general version. As we have said, the problematic property (the $T$-term) that is to be given a functional definition and reduction (the analogue of pain) is time. More precisely: it is duration, or the temporal metric. For in the `3+1’ framework of this paper, the unification of the temporal metric with the spatial metric in a Lorentz-signature metric---in physical terms: the relativity of simultaneity---will be `in the background’. In other words: although Lorentz-invariance will hold good (and be postulated or derived in diverse ways: reviewed in Section \ref{appetizer}), it will not be prominent.

The main idea of the temporal metric’s functional role, as it would be spelt out in a $T_t$, is that time is the measure, the parameterisation, of change. Stated as simply as that, the idea has a very long pedigree: from Aristotle through Descartes to McTaggart’s insistence that `time involves change’ \citet[p. 459]{Mctaggart}. But for us, the idea is of course more specific: time is to parameterise changes {\em of spatial geometry}, and its rate of ticking should in some sense mesh appropriately with the amounts of those changes. Thus the analogue of `being typically caused by tissue damage and typically causing distress and aversive behaviour’ will be along the lines: `being related in an appropriate way to spatial geometry---including to how it changes over time’. Here, the last phrase, `changes over time’, signals---as does the mental term `distress’, in the pain example---that in general, a $T$-term’s functional role involves other $T$-terms. But as discussed: using other $T$-terms does not give any problem of a logical circle. 

So much by way of $T_t$ (and thus premise i)). For the general version of $T_b$, i.e. the analogue of neurophysiology (and so premise ii)’s C-fibre firing), we suggest: the general features of a Hamiltonian theory of geometrodynamics. This will characterize time by: (i) its appearing in the denominator of the right-hand-side of equations $\{H, \varphi\}=\frac{d}{dt}\varphi$, governing the relevant fields $\varphi$; and thus implicitly by (ii) the symplectic flow of (or Poisson bracket with) the Hamiltonian in the left-hand-side.

We turn to the specific versions, one for each project. Of course, the details mentioned here (for example, the first project's deformation and constraint algebras) will become clearer in the following Sections.\\

\begin{itemize} 
 \item  (1): \citet{HKT}: Here one can take $T_t$ as the theory of (globally hyperbolic) Lorentzian spacetimes, and the functional role of time (or rather: of elapsed time, duration)  to be defined by the relative evolution of spacelike hypersurfaces, and its being required to obey the deformation algebra. One can take $T_b$ to be the Hamiltonian theory, whose constraint algebra realizes the deformation algebra. Thus in terms of Lewis’ notation, $\tau$ and $\rho$, introduced above: the $T_t$-term $\tau$ (the analogue of `being in pain’) equals: the generator of how spatial geometry changes over time, that reproduces the assumed deformation algebra of hypersurfaces.  And the $T_b$-term $\rho$ (the analogue of `C-fibre firing’) equals: the ADM Hamiltonian. In this way, Hojman et al.’s theorem, their uniqueness result, can be written as the Lewisian bridge-law: $\tau  = \rho$.
 
 \item (2): The programme of Schuller et al.:    Here it is clearest to express reduction in two stages, corresponding to the earlier work, reviewed in \citet{Schuller} and then the later work, revised with more modern techniques in \citet{Dull}. The first stage is the functionalist reduction of a theory about chrono-geometry to a theory about matter and radiation. The second stage develops the theory of gravity, considered as the evolution of a 3-geometry coupled to the first stage's dynamics of matter and radiation. The bridge between the stages is a set of consistency conditions, called `gravitational closure equations’: which Schuller et al. show can be satisfied. 
 
In the first stage, i.e. the functionalist reduction, one takes $T_t$ as a theory of  4-dimensional manifolds, {\em not} equipped with any metric, but equipped with matter-or-radiation fields whose equations of motion obey conditions about predictivity, good behaviour and being deduced from a diffeomorphism-invariant action functional. Then $T_b$ is a framework for generalized Lorentzian geometries: geometries that possess a generalized metric (not necessarily a contravariant tensor of rank two), by which we can identify causal cones, a future and past direction, and spacelike (in effect: Cauchy) hypersurfaces. Section \ref{sec:DGS} will give some details. The broad result is that again, from $T_t$ one could extract the functional role of the time parameter, along the lines that time is ``the dimension along which we make predictions’’: although, owing to the details, it would certainly be more cumbersome to state exactly than is the role extracted from Hojman et al.’s $T_t$.  
 
This leads to the consistency conditions, the gravitational closure equations. Thus in their second stage, Schuller et al. adopt the `3+1’ framework; and thereby connect to the technical apparatus of project (1), equating the commutation algebra between the Hamiltonian generators with that of the hypersurface deformation. As we announced in Section \ref{appetizer}: using this apparatus, Schuller et al. show that the consistency conditions are indeed satisfied by classical electromagnetism, with the resulting evolution of spatial geometry being described by the ADM Hamiltonian. 

Thus in this case at least, we can conclude that, with $\tau$  expressing the (admittedly cumbersome) functional role gestured at by the slogan that time is  ``the dimension along which we make predictions’’ and with  $\rho$ being again the ADM Hamiltonian: the functionalist reduction is again summed up by the Lewisian bridge law, $\tau  = \rho$.\\

 \item  (3): \citet{GomesShyam}:  In this project, one begins with only a 3-dimensional Riemannian manifold, and the general  (``philosophical'') idea above that its geometry changes. So although one assumes that time is the dimension of change, one does {\em not} assume that it forms a dimension of a spacetime manifold. Accordingly,  $T_t$ is a ``pre-relativistic'' theory about the time-evolution of causally unrelated quantities: quantities encoding facts about local geometry. Then time, understood as the measure of the local changes of these quantities (as elapsed time, duration), is functionally defined by the different local measures of change being required to commute in the appropriate way. (Details in Section \ref{sec:GS}.)  And $T_b$ is an avowedly Hamiltonian theory (as it was in projects (1) and (2)). It is a theory that implements, or realizes, the ideas of $T_t$ in a phase space formalism, subject to certain assumptions such as the Hamiltonian being at most second-order in momenta. 
 
 These conditions are rich enough to imply (with hard work!) that the Hamiltonian is the ADM one. So as for projects (1) and (2): with $\tau$  expressing the functional role of time given by the $T_t$ and with  $\rho$ once again the ADM Hamiltonian, the functionalist reduction can be summed up by the Lewisian bridge-law: $\tau  = \rho$.
  \end{itemize}

  \section{Realizing the temporal metric---from geometrodynamics}\label{sec:HKT}
We turn to project (1): `Geometrodynamics regained’  by  Hojman, Kuchar and Teitelboim, in  {\em Annals of Physics}, 1976; \citet{HKT}. For us this work is significant, not only as an important illustration of functionalist reduction in a spacetime theory; but also because (as we mentioned in Section \ref{appetizer}) some of its apparatus is also used in projects (2) and (3).  Following the `$T_t$ and $T_b$’ pattern of functionalist reduction, we present: in Section \ref{sec:hyper}, this project’s $T_t$, essentially the hypersurface deformation algebra; and in Section \ref{sec:constraint alg}, its $T_b$, essentially the constraint algebra. We state the uniqueness result for (the characterization of) the ADM Hamiltonian \citet{ADM} as Theorem \ref{theo:ADM_alg}. 

Of course, we will not display the entire proof here. That is not to our purpose; and anyway, its original exposition is very clear. But we will see, in the second half of Section \ref{sec:constraint alg}, that their result can be parsed as amounting to two main achievements.  The first, which we have already articulated, finds phase space functions whose algebra recovers the hypersurface commutation algebra. The second, by demanding that there is an algebra homomorphism between the constraints and the deformation vectors, guarantees that these phase space functions are actually constrained.  (Cf. Corollary \ref{cor:HKT} in Section \ref{sec:constraint alg}.) We will be interested in the second of these, since the apparatus involved will be deployed again in Sections  \ref{sec:DGS} and \ref{sec:GS}.

  \subsection{Hypersurface deformations}\label{sec:hyper}
      
 Hojman et al. assume a 4-dimensional Lorentzian manifold $M$, but adopt a `3+1' perspective. So they consider a 3-dimensional spatial manifold, which we will here call $\Sigma$. They assume it is compact and without boundary, and consider its embeddings as spacelike hypersurfaces of $M$.

  Specifically, suppose a foliation of spacetime by spacelike surfaces exists: $X: \mathbb{R} \times \Sigma\rightarrow M$. 
	 In this notation, $X$ takes a given $\sigma\in \Sigma$  to  an entire worldline (these are the points with the same spatial coordinate) and a given $t$ to an entire hypersurface.  Given a spacetime metric, $g_{\mu\nu}$, with signature $(-, +, +,+)$, at any point $X(\sigma, t')$, the tangent to this worldline can be decomposed  in  normal and tangential directions to the hypersurface: 
	 \be\label{eq:emb_dec}\dot X_{t'} :=\frac{d}{dt}_{|t=t'}X_t(\sigma)=N \mathbf{n}+\mathbf{s}=:(N, \mathbf{s})(x)=:(\dot X_{t'}{}_{\perp}, \dot {\mathbf{X}}_{t'}{}_{\parallel})(x)
	 \ee
  where $x=X_{t'}(\sigma)$ denotes the base spacetime point for the vector, $N$ is called \textit{the lapse}, $\mathbf{s}$ is the spatial vector field called \textit{the shift},\footnote{We have deviated slightly from the usual notation, in which the shift is also, somewhat confusingly from the 3+1 point of view, denoted by $\mathbf{N}$, or, in components, $N^i$. }; and in the last two equalities we represented this spacetime vector  in a doublet with its normal and parallel components to the hypersurface. \begin{figure}
		\begin{center}
			\includegraphics[scale=.5]{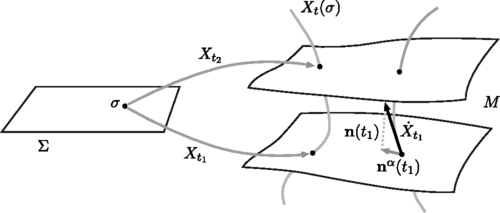}
			\caption{The foliation of spacetime, induced by a one-parameter family of embeddings of the model spatial manifold, $\Sigma$. The coordinates on $\Sigma$ are denoted by $\sigma^i$, and those on $M$ are denoted by $x^\mu$. (Taken from arxiv, with permission; from Dull et al, 2018.)  }
			\label{fig:foliation}
		\end{center}
	\end{figure}
  
  There is a representation of Diff$(M)$ onto the embedding maps, Emb$(\Sigma, M)$ (an infinite-dimensional space of mappings). That is, any spacetime diffeomorphism $d \in {\rm Diff}(M), d: p \mapsto d(p)$, defines a map, $\Delta_{d, \Sigma}$, on Emb$(\Sigma, M)$, as follows.
If $p = X(t, \sigma) \in M$, then $d$ defines another embedding (i.e. another member of Emb$(\Sigma, M)$), viz. $d \circ X$.
 Thus:
  $$\Delta_{d, \Sigma}: X \mapsto d \circ X $$
is a diffeomorphism of the infinite dimensional manifold Emb$(\Sigma, M)$.
 Correspondingly, a vector field $\xi$, being an infinitesimal generator of diffeomorphisms on $M$, defines an infinitesimal generator $D(\xi)$ of diffeomorphisms on Emb$(\Sigma, M)$; i.e. a vector field on Emb($\Sigma, M$); the map: $\xi \mapsto D(\xi)$ is a Lie algebra homomorphism.  We only have good control over the infinitesimal version of the embedding maps (and of the hypersurface deformation algebra below), and we will thus only work at that, infinitesimal, level.\footnote{ The reason is that it is  difficult to control the signature of the embedding after the application of a diffeomorphism. If we apply  a diffeomorphism to a   given spacelike embedding there is no guarantee  that the  hypersurface remains spacelike. Nonetheless,  under infinitesimal diffeomorphisms (i.e. vector fields) an originally spacelike embedding will remain spacelike, since being spacelike is an open condition: the normal $\mathbf{n}_x$  must be timelike at any point $x$, i.e. $g_x(\mathbf{n}_x,\mathbf{n}_x)<0$, which is an open condition. Thanks to Klaas Landsman for this point.  }

Thus proceeding at the infinitesimal level, on one side of this homomorphism, using the spacetime metric on $M$,  we can decompose  a  vector field  $\xi$  into components normal and parallel to the leaves of the hypersurface, $\xi=(\xi_\perp, \boldsymbol{\xi}_\parallel)$, with $\xi_\perp$ a scalar and $\boldsymbol{\xi}_\parallel$ a spatial vector field.

Then with $\xi^1, \xi^2$ vector fields, $[D(\xi^1), D(\xi^2)] = D([\xi^1,\xi^2])$ yields \citet{Teitelboim1973}:
  \begin{align}\label{eq:com_dec}
  [D(\xi^1_\perp, \boldsymbol{\xi}^1_\parallel), D(\xi^2_\perp, \boldsymbol{\xi}^2_\parallel)] &= D(\xi^3_\perp, \boldsymbol{\xi}^3_\parallel)
  \end{align}
  where
    \begin{align}
  \xi^3_\perp(\xi^1, \xi^2) &= \boldsymbol{\xi}^1_\parallel[\xi^2_\perp]- \boldsymbol{\xi}^2_\parallel[\xi^1_\perp]\label{vec_comm1}\\
   \boldsymbol{\xi}^3_\parallel(\xi^1, \xi^2) &= [\boldsymbol{\xi}^1_\parallel, \boldsymbol{\xi}^2_\parallel] - {\epsilon}\left(\xi^2_\perp h^{-1}({\rm d} \xi^1_\perp)-\xi^1_\perp h^{-1}({\rm d} \xi^2_\perp)\right)  \label{vec_comm2}
  \end{align}
    where $\epsilon=\pm 1$ is the spacetime signature---it is $-$ for Euclidean and $+$ for Lorentzian---and we have used coordinate-free notation (e.g.  $\boldsymbol{\xi}^1_\parallel[\xi^2_\perp]= {\xi^1}_\parallel^i \partial_i \xi^2_\perp$, and $\xi^2_\perp h^{-1}(\d \xi^1_\perp)=\xi^2_\perp h^{ij}\partial_i \xi^1_\perp$, where $h_{ij}$ is the pull-back of $g_{\mu\nu}$ to the hypersurface). 
      
  Essentially,  \eqref{eq:com_dec} gives the orthogonal decomposition of the commutator of (orthogonally decomposed) vector fields. It has a simple geometrical meaning: one is first  deforming the leaves of the foliation one way ($\xi^1$) and then another ($\xi^2$), and then comparing that resulting foliation with one in which the opposite ordering of deformations was applied. Suppose that the first vector field deforms the hypersurface $\Sigma$ into $\Sigma_1$, then deforming  that surface with $\xi^2$ we obtain $\Sigma_{12}$, i.e. $\Sigma\longrightarrow \Sigma_1\longrightarrow\Sigma_{12}$. The commutator $[D(\xi^1_\perp, \boldsymbol{\xi^1}_\parallel), D(\xi^2_\perp, \boldsymbol{\xi^2}_\parallel)]$ is the deformation map between $\Sigma_{12}$ and $\Sigma_{21}$. 
 
  The insight of Hojman, Kuchar and Teitelboim was to require that the  hypersurface evolution given by $D$ is matched by the geometrodynamical evolution given by the Hamiltonian. 

\subsection{Commutation algebras translated into constraint algebras}\label{sec:constraint alg}
The standard notation for the phase space variables---the metric on a space-like hypersurface and its conjugate momentum---is
$(h_{ab}(\sigma),\pi^{ab}(\sigma)),$ (the explicit dependence on space via
the label $\sigma$  reminds us that this is a pair of canonically
conjugate variables per space point in an infinite dimensional phase
space). So, at each point $\sigma\in \Sigma$ and before the imposition of constraints,  we have twelve phase space dimensions as both
the metric and its conjugate momentum are symmetric tensors.

We will denote the canonical Poisson brackets as $\{\cdot,\cdot\}$,
which, when applied to the fundamental phase space conjugate
variables $(h_{ab}(\sigma),\pi^{ab}(\sigma))$ yields 
$$
\{h_{ab}(\sigma),\pi^{cd}(\sigma')\}=\delta_{a}^{(c}\delta_{b}^{d)}\delta(\sigma-\sigma').
$$
 for $\sigma, \sigma'\in \Sigma$. Note that by $\delta(x-y)$ we really mean $\delta^{(3)}(\sigma-\sigma').$
Thus the action of the Poisson brackets on arbitrary phase space functions (omitting indices)
$F(g(\sigma),\pi(\sigma))$ and $P(g(\sigma'),\pi(\sigma'))$ is given by:
$$
\{F(h(\sigma),\pi(\sigma)),P(h(\sigma'),\pi(\sigma'))\}=\int\textrm{d}^{3}\sigma''\left(\frac{\delta F(\sigma)}{\delta h_{ab}(\sigma'')}\frac{\delta P(\sigma')}{\delta\pi^{ab}(\sigma'')}-\frac{\delta P(\sigma')}{\delta h_{ab}(\sigma'')}\frac{\delta F(\sigma)}{\delta\pi^{ab}(\sigma')}\right).
$$

To obtain their uniqueness result, Hojman et al.  assume  the total Hamiltonian for gravity is a sum: 
\be \label{eq:Ham_GR} \mathscr{H}(\xi_\perp, \boldsymbol{\xi}_\parallel)=\mathscr{H}_\perp(\xi_\perp)+\mathscr{H}_\parallel(\boldsymbol{\xi}_\parallel)
\ee
with 
\begin{align*}
\mathscr{H}_\parallel(\boldsymbol{\xi}_\parallel):=\int d^3\sigma\,{\xi}_\parallel^i(\sigma) \mathscr{H}_\parallel{}_i[h, \pi;\sigma)\quad \text{and}\quad
\mathscr{H}_\perp(\xi_\perp):=\int d^3\sigma\, \xi_\perp(\sigma)\mathscr{H}_\perp[h, \pi;\sigma). \label{dur_Ham}
\end{align*}
where we have used DeWitt's mixed index notation: $F[\pi, h;  \sigma)$ may depend  on the phase space data (and their derivatives) at each spatial point $\sigma$.  Hojman et al.   then require that the constraint algebra reproduces the deformation algebra:
\be\label{eq:HKT_alg}\left\{\mathscr{H}(\xi^1_\perp, \boldsymbol{\xi}^1_\parallel), \mathscr{H}(\xi^2_\perp, \boldsymbol{\xi}^2_\parallel)\right\}=\mathscr{H}(\xi^3_\perp, \boldsymbol{\xi}^3_\parallel) 
\ee
with $(\xi^3_\perp, \boldsymbol{\xi}^3_\parallel) $ obeying  \eqref{vec_comm1} and \eqref{vec_comm2}.

They also assume:\\

\noindent (1) The  generator of spatial diffeomorphisms is (with $\nabla$ the Levi-Civita covariant derivative associated to $h_{ij}$)
\be\label{Diff_gen} \mathscr{H}_\parallel{}_i[h, \pi;\sigma):=\left(h^{kj}\nabla_k\pi_{ij}\right)(\sigma) . \ee
That is: $$\left\{ h_{ij}(\sigma),\int d\sigma'\, \xi^\ell\left(\nabla^k\pi_{\ell k}\right)(\sigma')\right\}=\mathcal{L}_\xi h_{ij}(\sigma) \; ,$$
with the same holding for $\pi_{ij}$. They also assume \\

\noindent (2) The scalar generator $\mathscr{H}_\perp[h, \pi;\sigma)$ is such that: 
\be\label{eq:first_cond}
\left\{\int d^3\sigma\, \xi_\perp(\sigma)\mathscr{H}_\perp[h, \pi;\sigma) \; , h_{ij}(\sigma)\right\} \equiv \left\{\mathscr{H}_\perp(\xi_\perp), h_{ij}(\sigma)\right\} = 2 {\epsilon}(\xi_\perp K_{ij})(\sigma) \ee
where $K_{ij}$ is the extrinsic curvature of the embedded hypersurface in $M$. This is motivated by requiring $\mathscr{H}_\perp[h, \pi;\sigma)$ to generate diffeomorphisms normal to the leaves of the foliation.

With these assumptions, they then prove:

\begin{theo}[HKT]\label{theo:ADM_alg}Under assumptions \eqref{eq:HKT_alg},\eqref{Diff_gen}, and \eqref{eq:first_cond},   the unique choice of $\mathscr{H}_\perp[h, \pi;\sigma)$ is:
\be\label{Ham_ADM}
\mathscr{H}_\perp[h, \pi;\sigma) =\mathscr{H}_{\text{\tiny{ADM}} \, \perp}[h_{ab}, \pi^{ab};\sigma):=\left(a{ \epsilon}\frac{\pi^{ab}\pi_{ab}-\frac{1}{2}\pi^2}{\sqrt{h}}-a^{-1}(R-2\Lambda)\sqrt{h}\right)(\sigma)\ee
where $a$ and $\Lambda$ are  arbitrary constants.
\end{theo} 
This is  a remarkable uniqueness result. It entirely fixes the form of the generators of tangential and normal diffeomorphisms.

 But condition  \eqref{eq:first_cond} requires some justification, both technically and conceptually. First, technically: it requires the absence of terms containing derivatives of the momenta in the pure-normal part of the Hamiltonian. In fact, $\mathscr{H}_\perp[h, \pi;\sigma)$  is assumed to be a polynomial, ultralocal functional of $\pi^{ab}$.  Conceptually, the assumption has a physical meaning inherited  from spacetime intuitions. Namely, it encodes the fact that $\mathscr{H}_\perp[h, \pi;\sigma)$, or equivalently, $\mathscr{H}(\xi_\perp, 0)$, should represent an infinitesimal diffeomorphism normal to the leaves, i.e. a normal deformation in the spacetime embedding. Of course, this expectation is alien to the Hamiltonian vocabulary, and thus must be imported from the spacetime domain.\\

But arriving at \eqref{Ham_ADM} does not yet show that these generators are constrained, i.e. that they are generators of symmetries. 
The total Hamiltonian
of General Relativity is a sum of four constraints, the scalar Hamiltonian
constraint (one per point) and the vector diffeomorphism constraint (three per point). These constraints,
together with their Poisson algebra, encapsulate the initial value problem, the gauge symmetries and the evolution of the theory. But here we have so far only described $\mathscr{H}(\xi_\perp, \boldsymbol{\xi}_\parallel)$ as a symplectic generator of evolution, or of the  hypersurface  deformations. We have not yet discussed the reason for constraining these generators to be zero on-shell. 

This further restriction can still be extracted by imposing not only that the constraint algebra is the same as the deformation algebra, but that the algebra homomorphism works also at the level of generators of symplectic transformations. That is: the geometrodynamical evolution matches the geometrical deformation of surfaces. 

For that, we want to demand that:
\be\label{eq:main_meshing}\{\{\bullet, \mathscr{H}(\xi_1)\},\mathscr{H}(\xi_2) \}-\{\{\bullet, \mathscr{H}(\xi_2)\},\mathscr{H}(\xi_1) \}\approx\{\bullet, \mathscr{H}\}(\xi_3)\ee
where: $\xi_3$ is the commutator given in \eqref{vec_comm1} and \eqref{vec_comm2},  the last equality need  hold only on-shell,  $\bullet$ stands for any phase space functional, and we abbreviate, e.g. 
\begin{align}\left\{\bullet,\mathscr{H}_\perp\right\}(\xi_\perp)&=\int d^3\sigma\,\xi_\perp(\sigma) \left\{\bullet,\mathscr{H}_\perp[h, \pi;\sigma)\right\}\nonumber\\
\left\{\bullet,\mathscr{H}_\parallel\right\}(\boldsymbol{\xi}_\parallel)&=\int d^3\sigma\,\xi_\parallel{}^i(\sigma) \left\{\bullet,\mathscr{H}_\parallel{}_i[h, \pi;\sigma)\right\} .\label{eq:inside_comm}\end{align}

For phase-space-independent smearings, there is no difference according to where we place the Poisson brackets. But since the commutation algebra produces field-dependence, one must distinguish the two. What we mean by an evolution by a parameter $\xi$ is the latter, i.e. \eqref{eq:inside_comm}, where the parameter is placed outside the Poisson bracket. 

Using the Jacobi identity and \eqref{eq:HKT_alg} we have:
\be\label{eq:non_meshing1} \left\{\left\{ \bullet, \mathscr{H}(\xi^1_\perp, \boldsymbol{\xi}^1_\parallel)\right\}, \mathscr{H}(\xi^2_\perp, \boldsymbol{\xi}^2_\parallel)\right\}- \left\{\left\{ \bullet, \mathscr{H}(\xi^2_\perp, \boldsymbol{\xi}^2_\parallel)\right\}, \mathscr{H}(\xi^1_\perp, \boldsymbol{\xi}^1_\parallel)\right\}=\left\{\bullet, \mathscr{H}(\xi^3_\perp, \boldsymbol{\xi}^3_\parallel)\right\} \, .
\ee
But as can be seen from \eqref{vec_comm2}, 
$$ \boldsymbol{\xi}^3_\parallel = [\boldsymbol{\xi}^1_\parallel, \boldsymbol{\xi}^2_\parallel]-{\epsilon}\left(\xi^2_\perp h^{-1}(\d \xi^1_\perp)-\xi^1_\perp h^{-1}(\d \xi^2_\perp)\right) $$
clearly contains the metric $h_{ij}$; (equation \eqref{vec_comm1} does not).

 Therefore, from \eqref{eq:non_meshing1}, we would obtain: 
\be\label{eq:non_meshing2}\left\{\bullet,\mathscr{H}(\xi^3_\perp, \boldsymbol{\xi}^3_\parallel)\right\}=\left\{\bullet,\mathscr{H}\right\}(\xi^3_\perp, \boldsymbol{\xi}^3_\parallel)+\mathscr{H}_\parallel(\left\{\bullet,\boldsymbol{\xi}^3_\parallel\right\}) \, .\ee
From this, we can clearly see that the last term in \eqref{eq:non_meshing2} is the obstruction to obtaining \eqref{eq:main_meshing}. Note also that the last term is \textit{not} an infinitesimal diffeomorphism of $\bullet$. In fact, it will only  have an effect on the momenta of $\bullet$, since $\boldsymbol{\xi}^3_\parallel$ only contains the metric. Therefore, for the Hamiltonian evolution to \textit{mesh} in the right way, we must demand that, on-shell (in practice, `on-shell' means, implementing equality after  taking all functional derivatives, i.e. calculating all the Poisson brackets):
\be \mathscr{H}_\parallel(\boldsymbol{\xi}_\parallel)\approx 0, \,\, \forall \boldsymbol{\xi}_\parallel\in C^\infty(T\Sigma)\qquad\text{or equivalently}\qquad \mathscr{H}^i_\parallel[h, \pi;\sigma)\approx 0 \; \;  ;
\label{eq:diff_ctraint}\ee
where on-shell equality is denoted by a $\approx$. But to be conserved, this condition will also imply the vanishing of $\mathscr{H}_\perp$:
$$\left\{\mathscr{H}_\perp(\xi^1_\perp), \mathscr{H}_\parallel(\boldsymbol{\xi}^2_\parallel)\right\}=\mathscr{H}_\perp(\boldsymbol{\xi}^2_\parallel[\xi^1_\perp])\approx 0 , \,\, \forall \boldsymbol{\xi}^2_\parallel\in C^\infty(T\Sigma),\forall \xi^1_\perp\in C^\infty(\Sigma) \, .
$$

Therefore, the final condition required to  obtain general relativity in its ADM form from Theorem \ref{Ham_ADM}  is \eqref{eq:main_meshing}; or, assuming  \eqref{eq:HKT_alg} (and using the Jacobi identity), the required condition is: 
\be\label{eq:meshing_final} \left\{ \bullet, \mathscr{H}(\xi^3_\perp, \boldsymbol{\xi}^3_\parallel)\right\}\approx\left\{\bullet,\mathscr{H}\right\}(\xi^3_\perp, \boldsymbol{\xi}^3_\parallel)
\ee
where $\xi$ may have phase-space-dependence (in the example above, $\xi=\xi^3$ had a specific dependence on the metric, as a result of the commutation algebra). 

Mathematically, this condition is necessary to ensure that we can interpret the mapping of the deformation vectors into a lapse and shift as a  homomorphism from the algebraic structure
of hypersurface deformations into the derivations of phase-space functions, i.e. as Hamiltonian evolution. 
That is:, we want to guarantee that the successive evolution by a pair of parameters  stays within the same phase-space solution curve, irrespective of their ordering.  Hojman et al. interpret this demand as  encoding the ``path-independence'' of the evolution of geometrical quantities---the irrelevance of the intermediate foliations \citet[Sec. 5]{HKT}.

Thus we can now state:
\begin{cor}\label{cor:HKT} Under assumptions  \eqref{eq:main_meshing},\eqref{Diff_gen}, and \eqref{eq:first_cond} (that is, replacing \eqref{eq:HKT_alg} by the stronger \eqref{eq:main_meshing}) we obtain ADM Hamiltonian geometrodynamics.  
\end{cor}
Once one has recovered foliation invariance, and all that comes with it, the closure of the brackets translates back in covariant language to the Bianchi identities. The geometrodynamical coupling of matter then follows suit: it corresponds to the Legendre transform of a generally covariant matter theory coupled to a general relativistic spacetime.\footnote{Nonetheless, the consistent propagation of matter can also be obtained in the canonically 3+1 language, as above. }

Note that, since it is only the normal components of the two diffeomorphisms that form the obstruction in \eqref{eq:non_meshing2}, we can state path-independence employing only equivalence through spatial diffeomorphisms.  That is,  for `pure' time evolution, setting the parallel components of $\xi$ to zero, one must demand only that  \textit{the geometry} of $\Sigma_{12}$ is the same as that of $\Sigma_{21}$ (i.e. that they differ by the action of a spatial diffeomorphism). Of all of the principles of  Hojman et al's  reconstruction, only this weaker `path-independence', or meshing condition, will be  imported into the non-spacetime-based theorems of Section \ref{sec:GS}.

\section{Realizing the temporal metric---from matter without geometry }\label{sec:DGS}

We turn to project (2): the programme of Schuller and his collaborators \citet{Schuller, Dull} to recover chrono-geometry, specifically, geometrodynamics, from postulates about the physics  of (massive or massless) matter fields defined on a 4-dimensional manifold, but without assuming {\em ab initio} a Lorentzian metric. This is clearly a programme that is both ambitious and avowedly `relationist'. So it deserves the attention of philosophers of physics, especially those sympathetic to relationism (e.g. the dynamical approach of  \citet{Brown_book}).\footnote{But so far as we know, the only philosophical discussion of it is \citet{Menon_lor}, which we commend.} All the more so, since the programme has been developed in detail and has achieved a great deal. But our discussion of it must be limited: we will confine ourselves to showing how the programme illustrates, not just reduction, but our preferred functionalist reduction.

As we said at the end of Section \ref{3illustr}, it is best to think of the reduction in two stages. We give an overview of both stages in Section \ref{sec:2stage}, then give their details (Sections \ref {sec:closure} and \ref {sec:DGS_HKT} respectively). Finally, in Section \ref{sec:counter}, we will sketch an example showing the programme’s dependence on assuming diffeomorphism invariance  (as we announced in Section \ref {appetizer}).

\subsection{Overview of the two stages of reduction}\label{sec:2stage}
The broad idea is that the various familiar roles that the Lorentzian metric plays in relativistic physics are to be replaced by, indeed reduced to, conditions on matter and radiation: conditions that are less problematic from a relationist viewpoint. However, Schuller et al. do assume from the start: a 4-dimensional manifold, and that the equations of motion for matter and radiation are deduced from a fully diffeomorphism-invariant action functional. (So this precludes certain theories that take a fundamentally 3+1 perspective---theories such as shape dynamics \citet{SD_first, SD_linking} and Horava-Lifschitz gravity \citet{Horava}.) 

Their first stage then consists of adding some reasonable assumptions about these equations (details below), and deducing that spacetime has a causal structure and, indeed, a generalized geometry. More precisely: they deduce that there are well-defined future and past causal cones (limiting the speed of causal propagation) at each point, and this yields  definitions of timelike trajectories and spatial---or rather, Cauchy---hypersurfaces, as well as mass hyperbolas. We will call this a {\em kinematical geometry}.\footnote{Although this is our label, not these authors', it is  prompted by some of their phrases like `kinematical meaning' in the next quotation.}  Then the second stage requires that both the matter and the geometry of a spacetime canonically evolve together, starting and ending on shared Cauchy surfaces, and independently of the intermediate foliation. This requirement means a set of consistency conditions, called `gravitational closure equations', must be satisfied. Schuller et al. then deploy  a version of the apparatus of Hojman et al, i.e. project (1) of Section \ref{sec:HKT}---suitably generalized to their  kinematical geometries---to show that  classical electromagnetism  satisfies these consistency conditions. Thus they recover orthodox geometrodynamics, i.e. the orthodox general relativistic canonical co-evolution of the Maxwell field and spatial geometry---without having assumed {\em ab initio} a Lorentzian metric!  Clearly, this is a major achievement, even though it is restricted to classical electromagnetism; (so that in philosophers' jargon, it is a local reduction).

These two stages are  summarised successively in the introduction to \citet{Dull}; and this is followed by a remark which reminds us of the dynamical approach of  \citet{Brown_book}: 
\begin{quote}
``[We first prescribe] the matter dynamics on a spacetime. The dynamics of the underpinning spacetime geometry
are then shown to follow from the matter dynamics, essentially by a sufficiently precise requirement of common canonical evolution ...  
 the matter dynamics crucially determines the kinematical meaning of their geometric background [i.e. first stage], and it is precisely this
information that directly funnels into the structure of the gravitational dynamics [i.e. second stage]. ... And [this is] not a new perspective either, considering that it was the dynamics of matter, namely the classical electromagnetic field, that led Einstein to the identification and kinematical interpretation of Lorentzian geometries and finally the field equations for their dynamics [cf. the dynamical approach].'' (2018: Introduction, p. 084036-1)
\end{quote}

We end this Section with a couple of details and quotations, about the first stage. At the start, one assumes three conditions about the matter: (technically, conditions on the principal polynomial of the corresponding field equations). In the authors' words \citet{Dull}:
\begin{quote} ``Classically [these three conditions] correspond, in
turn, to: the existence of an initial value formulation for the matter field equations; a one-to-one relation between momenta and velocities of massless particles; the requirement of an
observer-independent definition of positive particle energy. It is interesting to note that if one
insists on the matter field equations being canonically quantizable, these three properties are
directly implied.''   (2018: Introduction, p. 084036-2)
\end{quote}

Moreover, one assumes that the equations of motion are obtained  from extremising a diffeomorphism-invariant action functional $S_{\text{\tiny matter}}$.  As we will see, this second assumption is central. For one thing,  it means assuming that there is some spacetime density or volume-form, i.e. some way to count infinitesimal volume---which will be a ``seed'' for metrical relations. And one assumes that additional to the matter degrees of freedom, there is a tensor $G$ on the spacetime, that is constrained only by how it enters the action---through the volume-form for example, and perhaps by how it contracts indices of derivatives or tensor indices of the matter fields. (So nothing about the signature, or tensor valence, of $G$ is assumed: it could be given by a symplectic-form, or by a four-valence tensor contracting area-elements, etc.) 

$G$  is called a  {\em geometric tensor}; though in the absence of the other assumptions, it will in general not define a causal structure or other geometrical notions.  (So while `G’ naturally stands for `geometry’, one might also think of it as standing for `gadget’!) But the three conditions on matter fields are then translated into conditions on  dispersion relations which, in their turn, are  encoded in a totally symmetric contravariant even-rank tensor field, satisfying three simple algebraic conditions. One can then identify much of the structure, such as a causal (cone) structure, usually associated with there being a Lorentzian metric---even though one has not assumed such a metric. In  Schuller's words:
\begin{quote}
``These physically inevitable properties single-handedly ensure that the entire kinematical apparatus familiar
from physics on a Lorentzian manifold is defined in precisely the same way for any spacetime;
causality, in particular, is perfectly compatible with superluminal propagation in spacetimes,
but one only learns this from a subtle interplay of convex analysis, real algebraic geometry
and the modern theory of partial differential equations." \citet[p.2 ]{Schuller}
 \end{quote}

Then it turns out that whatever the geometric tensor $G$, and whatever the matter fields (as long as they satisfy the assumed conditions),  a set of consistency conditions on the deduced spacetime structure must be satisfied---leading  into the second stage.\footnote{In general terms, it is unsurprising that there should be such consistency conditions. Recall that in the covariant framework,  Lovelock's theorem \citet{Lovelock} severely restricts the possible Lagrangian densities for the gravitational part of the action. Agreed: that theorem assumes that the ``felt'' spacetime metric is the standard Lorentzian metric $g_{\mu\nu}$: without that assumption, the theorem does not apply, and one must  seek more general theorems constraining the  gravitational part of the action. But instead of seeking a generalized version of Lovelock's theorem, Sch\"uller et al. adopt the alternative, canonical framework, and so can invoke known consistency requirements, drawing on Section \ref{sec:HKT}.}

\subsection{Kinematical geometry from matter dynamics}\label{sec:closure}

As we have admitted, a detailed account of this programme  would go far beyond this paper: a summary of the general ideas must suffice. The starting assumption is the action for matter  or radiation, described by a tensorial field $B$ on the spacetime manifold $M$, which is ultralocally coupled to a ``geometric'' tensor field, $G$. $G$ is of arbitrary tensor valence. It is needed at least to form scalar densities  of weight one (i.e. ensuring diffeomorphism invariance for $S_{\text{\tiny matter}}$), but it can also  be used to contract indices of the same valence:\footnote{Here we are supposing a single geometric tensor. In \citet{Dull}, it is shown how to incorporate multiple $G$'s. }
\be\label{eq:Matter_lagrangian}
S_{\text{\tiny matter}}[B; G)=\int d^4 x \mathcal{L}_{\text{\tiny matter}}(B,  \partial^{(K)} B; G(x))
\ee
where $\partial^KB$ denotes dependence on an arbitrary but finite number of derivatives of $B$, i.e.  from $\partial_i B$ to $\partial_{i_1}\cdots\partial_{i_F} B$.  From this action, it is assumed that the equations of motion for $B$ are of the form: 
\be\label{eq:eom}
Q^{i_1\cdots \i_n}_{\mathcal{AB}}(G(x))(\partial_{i_1}\cdots\partial_{i_F} B^{\mathcal{A}})(x)+\mathcal{O}(\partial^{F-1}B)=0 \; .
\ee
In this equation, $\mathcal{A,B}$ denote the tensor indices of the matter field.  $Q^{i_1\cdots \i_n}_{\mathcal{AB}}(G(x))$ is a tensor density of weight one by itself, functionally dependent on the geometric tensor. Despite the appearance of partial---not covariant---derivatives, by ignoring the lower order terms (where the non-covariant terms would appear), \eqref{eq:eom}  is made appropriately covariant. It is important that the equations of motion are linear in the field $B$, for otherwise one cannot isolate the implications of the dynamics for the causal structure.\footnote{The terminology `probing matter', as introduced in \citet[p. 12-13]{Schuller}, signifies matter with linear equations of motion. As he describes it, with non-linear equations ``it would often be impossible to disentangle the properties of the matter  field from properties of the underlying geometry''.}

While the deduction of the ``felt'' spacetime properties from $Q$  is not at all straightforward, we can  illustrate the gist of it with a simple example: consider the Wentzel-Kramers-Brillouin expansion of the matter field: 
$$ B^{\mathcal{A}}(x)=\exp{ (iS(x)/\lambda)} (b^{\mathcal{A}}+\mathcal{O}(\lambda)) \; .
$$
Then, from applying the equations of motion, to lowest order we obtain: 
$$ Q^{i_1\cdots \i_n}_{\mathcal{AB}}(G(x))k_{i_1}\cdots k_{i_F} b^{\mathcal{A}}=0
$$
which precisely captures the infinite frequency limit of the fields. Of course, these equations need not define anything like the causal cones of standard Lorentzian metrics: for starters, here the $k_i$'s are momenta, and must be (Legendre) transformed into vectors to constitute elements of spacetime. It is part of the merit of  \citet{Schuller} to show that  in fact they do define causal cones---when  \eqref{eq:eom} satisfy the necessary conditions.

Generalizing the manipulation of $Q$ above,  one defines the \textit{principal polynomial}, which carries most of the dynamical significance of the equations of motion:\footnote{More care is needed in the presence of gauge symmetries, but that case too is encompassed by the framework of \citet{Dull}.}
$$ P(x,k)=\pm\rho \det_{A,B}\left(Q^{i_1\cdots \i_n}_{\mathcal{AB}}(G(x))k_{i_1}\cdots k_{i_F}\right)
$$
If the (linear, partial differential) equations of motion have a well-posed initial value-problem, then $P(x,k)$ defines a hyperbolic homogeneous polynomial in each cotangent space, $T_x^*M$, where $\rho$ is a scalar density obtained from $G$. Here, the polynomial  is called `hyperbolic'
if there is a covector $k\in T^*_xM$ such that $P(x,k)\neq 0$ and the equation $P(x, p+\lambda k)=0$ has only real solutions for $\lambda$, for any further covector $p \in T^*_xM$. 

This way of describing $P$ leads directly to the definition of hyperbolicity cones, which are the seeds for defining the causal structure of the theory directly from the equations of motion, and which in fact generalize the familiar causal cones in Lorentzian geometry.\footnote{For higher order polynomials, one gets  an even number of distinct hyperbolicity cones, but not necessarily two, as in Lorentzian geometry. Importantly, this discrepancy does not block general dynamical considerations; and in many cases, there are work-arounds. For instance, for reducible polynomials (i.e. forming many null directions, since $P(x)=P_1(x)\cdots P_n(x)$), one obtains the overall hyperbolicity cone by intersection.} Namely, for each such $k$,  there is always an open and convex cone $C_x(P; k)$   that contains all hyperbolic covectors that lie together with $k$ in one connected set. It is a cone not in the geometric sense {\em per se}, but algebraically: for every $k'\in C_x(P; k)$, we have $ C_x(P; k')= C_x(P; k)$, and for $k', k''\in  C(P, k)$,   $\lambda k'\in C(P, k)$ and $k'+k''\in  C(P, k)$. To render it geometrical, one must convert cotangent vectors  to vectors.  

Another distinguishing feature (which we will not delve into), is that in the generalized case, the dual polynomial (e.g. defining the causal structure through vectors instead of co-vectors)  may not even be  completely well-defined.\footnote{A dual principal polynomial to $P$ at $x$ is any principal polynomial $P_x^\#$ that acts on the tangent space and annihilates tangents to the bases of null cones of $P$. That is,   let the causal cones (in cotangent space) be defined as $N_x:= \{k\in T^*_xM : P(x,k)=0\}$, with smooth subcones defined by  $N^{\text{\tiny smooth}}_x:= \{k\in N_x : \text{grad}_xP(x,k)\neq 0\}$. Then $P_x^\#:T_x M\rightarrow \bb R$ is  defined by $P_x^\#(\text{grad}_x(N^{\text{\tiny smooth}}_x))=0$. Then $P$ is bi-hyperbolic if both $P$ and $P^\#$ are hyperbolic; and similarly: for $C_x$, one can define $C_x^\#$.} In fact, finding the dual polynomial is instrumental for a proper Legendre transformation of the matter-induced dynamical structure (such as `observer worldlines'). A  Legendre map $L$ is essentially the spacetime gradient (at the base point) of $\ln P(x,k)$, that is: $L_x: C_x\rightarrow T_xM \,:\, k \mapsto -\text{grad}_x(\ln P)(x, k)$. It is well-defined and exists when $P$ is bi-hyperbolic and energy-distinguishing. Such a Legendre map will take elements of the hyperbolicity cone $C_x$ to elements of $T_xM$. In this case,   one can move from dynamical structures in the cotangent spaces $T^*M$ to structures in $TM$ and, eventually, to structures in spacetime $M$. For instance, for `an observer' curve, whose tangent $\dot\gamma$ is always within the observer cone $C_x^\#$, the vector space $V_x(\dot\gamma):=\{X\in T_xM : L^{-1}(\dot \gamma)(X)=0\}$ defines the purely spatial directions as seen by such an observer.\footnote{Although Schuller in \citet{Schuller} uses different maps for the null cone and the time-like trajectories---the Gauss map and the Legendre map, respectively---this is not strictly necessary to build a geometry in the sense above, and the distinction is indeed dropped from the later papers in the series. }

Proceeding in this manner---though we should note that this construction does not directly provide a Lorentzian spacetime metric---one is ultimately able with hard labour to use the dynamics to identify spatial hypersurfaces, temporal directions, momenta, observer trajectories, observer frames, particle energies, etc.

In sum, the great merit of this programme \citet{Schuller, Giesel_Schuller, Dull}  is that it builds a causal spacetime notion directly from the dynamics of matter. But it is not true that `anything goes'. For even after satisfying the conditions on the matter equations, consistency conditions emerge once one couples the dynamics of matter to the auxiliary geometrical structures, $G$---leading us to the second stage.

   \subsection{Geometrodynamics from kinematical geometry}\label{sec:DGS_HKT}
   
   The equations of motion also bequeath    spatial geometric structure to the spacelike surfaces and temporal geometric structure to the remaining, causal directions. More precisely:  the existence and uniqueness of the way to associate normal directions along a hypersurface with its canonical normal co-directions by means of a Legendre map requires all three algebraic properties: the hyperbolicity, time-orientability and energy-distinguishing properties.
   
But once we have such surfaces and their orthogonal directions, one can  deform the hypersurfaces, either  orthogonally to the surfaces, or parallel to them. Applying such deformations in succession, one can construct a purely geometric \textit{hypersurface deformation algebra}, as seen in section \ref{sec:HKT}, and equation \eqref{eq:com_dec} (see \citet{Teitelboim1973}).

   With these algebras in hand, we can apply a generalized form of the Hojman-Kuchar-Teitelboim (HKT) procedure \citet{HKT}, also described  in section \ref{sec:HKT}, by which one defines Hamiltonian generators in the gravitational phase space. These generators are designed to recover the action of the hypersurface deformations, and are thus thought, in some sense, to generate spacetime diffeomorphisms. 
   
  In sum, the procedure consists in the following steps. First, we  use the dynamically-induced causal structure to define not only the spatial hypersurfaces, but also its associated normal vectors in the time direction. Then, one obtains a hypersurface deformation algebra by decomposing general spacetime vector fields in terms of this foliation. This need not match the standard metric spacetime hypersurface commutation algebra; the difference is due to the kinematical information in the construction of the hypersurfaces. 
  Thus, upon computing the hypersurface deformation algebra, only the pure time-time component can differ from \eqref{vec_comm1} and \eqref{vec_comm2}: 
\begin{align}
[D_\perp(\xi^1_\perp), D_\perp(\xi^2_\perp) ]&=\mathscr{H}_\parallel\left(K[Q]^{ij}(\xi^1_\perp\partial_i \xi^2_\perp- \xi^2_\perp\partial_i \xi^1_\perp)\right)\label{1}
\end{align}
The tensor $K[Q]^{ij}$ is defined implicitly by $Q$ (see eq 45 and 48 in \citet{Dull}). It is only in this dynamical relation that the equations of motion make an  appearance. 

 In parallel, using the `gadget' geometrical tensor $G$, one can find the induced geometrical structure on these spatial hypersurfaces---call these $\phi(\sigma)$ for $\sigma\in \Sigma$, the spatial hypersurface---and a corresponding phase space for these quantities---labelling as $\pi(\sigma)$ the associated conjugate momenta to $\phi(\sigma)$ (the actual definition is also laborious, and involves again the principal polynomial). Finally,  one requires that the hypersurface deformation generators be emulated in the phase space, by respective Hamiltonian generators. That is, equating the commutation algebra between the Hamiltonian generators with that of the hypersurface deformation, one obtains further consistency conditions---this time on the geometrodynamical part of the action. In a nut-shell, the matter dynamics in this way restricts the geometrodynamics of any model embedded in a spacetime manifold.\\

Finally, two last conditions, in the spirit of Hojman et al,,  are then assumed in order to obtain the compatible gravitational dynamics. They are called in \citet{Dull}:  (i) ``local phase space avatars of deformation operators''---corresponding to a matching of the Hamiltonian constraint algebra and the ``felt geometry'' deformation algebra, and (ii) ``spacetime diffeomorphism invariance'' ---corresponding to \eqref{eq:main_meshing}.

The two conditions together require the existence of  constraint functions on phase space: 
\be \mathscr{H}_\perp(\xi_\perp)=\int d^3 \sigma\,\xi_\perp(\sigma) \mathcal{H}_\perp (\phi, \pi)(\sigma)\approx 0\qquad \text{and}\qquad\mathscr{H}_\parallel(\boldsymbol{\xi}_\parallel):=\int_\Sigma d^3 \sigma\, \xi^i_\parallel(\sigma)\mathcal{H}_i^\parallel (\phi, \pi)(\sigma)\approx 0\ee
satisfying the same algebra as in \eqref{1} and \eqref{vec_comm1}. As before, the demands of (i) spacetime embeddability and (ii) homomorphism of deformations to symplectic generators imply that the total Hamiltonian for gravity  be a fully constrained sum: 
\be \label{eq:Ham_GR} \mathscr{H}_Q(\xi_\perp, \boldsymbol\xi_\parallel)=\mathscr{H}_{Q\perp}(\xi_\perp)+\mathscr{H}_\parallel(\boldsymbol\xi_\parallel)
\ee
where we have inserted a subscript $Q$ to remind ourselves that the Hamiltonian could in principle be different from the standard ADM Hamiltonian, depending on the matter dynamics. 

Thus we can finally state:
\begin{theo}[Schuller et al]
For any diffeomorphism-invariant matter action whose integrand depends locally on some tensorial matter field $B$ and ultralocally on a geometric background described by some tensor field $G$ of arbitrary valence, 
\be S_{\text{\tiny{matter}}}[B; G) =\int \d^4 x L_{\text{\tiny{matter}}}(B(x), \pp B(x), \dots, \pp_{\text{\tiny{finite}}}B(x); G(x)) \ee 
and that satisfies the three matter conditions (the existence of an initial value formulation for the matter field equations; a one-to-one relation between momenta and velocities of massless particles; the existence of an observer-independent definition of positive particle energy):---\\
\indent \indent one can calculate four geometry-dependent coefficients that enter the equations that guarantee that the geometrodynamical commutation algebra emulates the hypersurface deformation algebra (i.e. the gravitational closure equations), which can be posed as a countable set of linear homogeneous partial
differential equations. Then a solution of these equations, in turn, provides a consistent action coupling the matter to the geometry:
\be S_{\text{\tiny{closed}}}[B; G)= S_{\text{\tiny{matter}}}[B; G)+\int \d^4 x L_{\text{\tiny{geometry}}}(G(x), \pp G(x), \dots, \pp_{\text{\tiny{finite}}}G(x))
\ee 
\end{theo}

Note that the theorem does not state the existence of such a solution for any given $B$ and $G$ and equations of motion. Apart from known  examples (such as classical electromagnetism)  
which recover GR: so far as we know, no other geometrodynamical theory has been found by explicitly solving the gravitational closure equations. 

Ideally, the uniqueness theorem we seek would answer the following question: allowing for different matter dynamics in the space of possible derived geometries, is the dynamics of  the standard model unique in singling out a meaningful dynamical spacetime? Or are there other families of dynamical fields which would imply different types of (generalized) geometries? Alas, no such theorem is known.   In this sense, the programme does not as yet inform us whether chrono-geometry as embodied by a Lorentzian spacetime can or cannot be separated from the dynamics of matter. All we know so far is that for the dynamics of matter that we have, e.g. classical electromagnetism, Lorentzian geometry is what we get. But so to speak: we already knew that. In that sense, unfortunately, the programme does not yet help settle whether matter dynamics or chronogeometry should have explanatory priority over the other. But setting aside this limitation, the programme has the great merit of providing a framework  in which the matter dynamics poses strict consistency conditions on any gravitational theory which is to accompany it; and  in the future, more powerful statements might well be extracted from this framework.

\subsection{A counter-example without spacetime diffeomorphism invariance}\label{sec:counter}

 We have already seen that this programme’s assumptions include the action functional being diffeomorphism invariant. In this final Subsection, we bring out the programme’s need for this assumption, by pointing to an example that in its absence, exhibits electromagnetism with the standard coupling to a Lorentzian metric, but also a {\em non}-orthodox geometrodynamics, i.e. without an Einstein spacetime, or the strong equivalence principle.
 
This counter-example  \citet[p. 317]{Gomes2017}  is predicated on a symmetry principle distinct from  spacetime diffeomorphism invariance: namely,  spatial conformal and diffeomorphism invariance. With this principle, one can stipulate an action that is radically different than the orthodox Einstein-Hilbert one,\footnote{Without a reference density, one can build a conformally invariant 3+1 action by using the square root of the Cotton tensor contracted with itself, cf. \citet[p. 314]{Gomes2017}. Such an action has a completely different structure than the Einstein-Hilbert one, even restricting to the simplest second-order in time-derivatives. } and one can still couple spin-1 fields and gravity within a consistent dynamics.  And when the conformal curvature is small, the equations of motion obtained by the action up to second order in derivatives recover precisely the Maxwell dynamical equations in temporal gauge. Nonetheless, the spacetime obtained from the dynamics need not satisfy the Einstein equations.\footnote{The consistency conditions between the geometric tensor $G$ and geometrodynamics found in \citet{Schuller}, ensue only if one further  assumes the matter dynamics generates  \textit{local} hypersurface deformations. This is not at all necessary: the matter dynamics may define the translation of one hypersurface to the next  in a global manner. }

Thus, without assuming spacetime diffeomorphism invariance from the outset, the dynamical equations---even Maxwell's---seem insufficient to determine the  orthodox spacetime geometry,  i.e. orthodox geometrodynamics.

\section{Realizing the temporal metric---from only spatial geometry}\label{sec:GS}

Let us introduce project (3) by sketching how it relates to projects (1) and (2). More precisely, we will sketch how those projects prompt an expectation about the role of spacetime---which project (3) defeats.

In section \ref{sec:HKT},  we saw that  \citet{HKT} found a direct path from spacetime properties to a Hamiltonian formalism: namely by emulating the  algebra of embedded surfaces as a constraint algebra in the Hamiltonian formalism. Then in Section \ref{sec:DGS}, we reported on work \citet{Schuller} tying (certain types of) matter dynamics to the dynamics of spacetime.  We also pointed to obstacles to a unique specification of standard gravitational dynamics, once one no longer assumes spacetime and its associated requirement of diffeomorphism invariance of the action from which the dynamics are to be derived. 

So in the light of these projects, one might expect that a 4-dimensional spacetime needs to be assumed in order to arrive at a consistent dynamics of gravity and matter. That would be a blow to relational approaches such as shape dynamics: which, as we will see in Section \ref{sec:payoff}, aims to base its ontology not on spacetime, but on conformal superspace. More specifically, it would suggest that shape dynamics must somehow  ``piggy-back'' on a spacetime framework.  

But this is not so. In this Section, we will show that weaker requirements---which do not assume spacetime or spacetime diffeomorphism invariance, at least in its usual form---suffice to deduce general relativistic dynamics. That is:  without presupposing spacetime, we will deduce precisely the Hamiltonian of general relativity, i.e. of the form \eqref{Ham_ADM}, for the appropriate scalar and vector constraints  of the ADM formalism. In short: while Hojman et al. found a path from spacetime to a Hamiltonian formalism, here we will describe a path in the opposite direction.

\subsection{Constraints on duration}\label{constrduratn}

We begin with the `top theory' $T_t$, a theory of Riemannian geometry on a 3-manifold.  More precisely, we envisage a configuration space, $\Phi$,  whose elements $\varphi$ are representations of the instantaneous state of the Universe. By `instantaneous' here, we do not mean any particular spacelike embedding of a hypersurface into a spacetime---there is no spacetime yet! We mean only that  the fields at different points are dynamically unrelated.  Thus we also assume that the state of the Universe can depend on location. That is, we assume that the theory is not altogether topological. Since we are dealing with field theories, this means that the theories in question will have an infinite number of degrees of freedom (at least one per point of space). 
 
  In this picture, since change can be measured locally, it seems natural to also associate duration with change {\em locally}. But duration is required to  mesh in the right way; we will require that the end result after evolution is independent of our choice of intermediate states. 
  That is, a final state is independent of the order in which two different changes are successively applied to an initial state. At this point, we follow a slightly altered version of  Hojman et al.'s demand of ``path-independence". Their demand was that one falls on the same integral curve of the Hamiltonian, irrespectively of the order of evolution. Ours will be a version specialized to pure time evolution, and applying an equivalence relation based solely on the `instantaneous' field content of the theory.

  More precisely, denoting two  evolutions for $\varphi$, that may be distinct at distinct locations,  by $\Delta^1_\perp\varphi$ and $\Delta^2_\perp\varphi$, the consecutive evolution of the two should be independent of the intermediate state. Thus we demand: $\Delta^{12}_\perp\varphi=\Delta^{21}_\perp\varphi$. (We could omit the $\perp$, since here we do not have embedded surfaces; but it is an intuitive label.)

\subsection{Constraints on evolution}\label{constrevoln}
  
  Now we move to sketching the formulation of a Hamiltonian geometrodynamics, as our $T_b$. As announced, the main result (Theorem \ref{theo:rigidity} below) will be that in this theory, the ADM Hamiltonian \eqref{Ham_ADM} is the unique realizer of the functional role of $\Delta_\perp$.   Since we seek a Hamiltonian theory, the `instantaneous state' of the Universe will be a phase space doublet, $(h_{ab}, \pi^{ab})$. And invoking the idea of spatial geometry, we require that the theory  should  depend on the doublet $(h_{ab}, \pi^{ab})$ only up to diffeomorphisms. That is, the theory should be invariant under spatial diffeomorphisms.

Our definition of the Hamiltonian generators of change will, therefore, take the form of constraints that the momenta and the metric have to satisfy.  But here, we will not demand that the symplectic generators of the diffeomorphisms reproduce the algebra of hypersurface deformations---for in the present picture, we have neither spacetime diffeomorphisms nor hypersurfaces. We will not even demand Section \ref{sec:HKT}'s full meshing relation, \eqref{eq:main_meshing}: we will only need a similar relation for the pure time evolution, that is, the equivalent of $\Delta^{12}_\perp\varphi=\Delta^{21}_\perp\varphi$, up to the equivalence relation of the theory, namely, spatial diffeomorphisms. This equivalent is \eqref{eq:GS_geomequiv} below.
 
 We will also assume that, whatever constraint we find for time evolution, it must be a scalar. This is based on the demand on $T_t$ that the theory contain local degrees of freedom. The requirement  is not mentioned by Hojman et al. because the evolution normal to a hypersurface has only one direction (per point), and is thus already scalar. Here, without the aid of the spacetime picture, we cannot appeal to that argument. 
 
 Here, the argument for a scalar Hamiltonian  (i.e. that the Hamiltonian can depend at most on a scalar Lagrange multiplier) just counts degrees of freedom. Let us rehearse how this goes. In three dimensions,  $h_{ij}$ has six degrees of freedom; diffeomorphisms---being generated by vector fields (Lagrange multipliers with three indices per space point)---take away three of these. Since  the Hamiltonian must preserve spatial diffeomorphism symmetry, as a constraint it should be first class with respect to the generator of spatial diffeomorphisms. Therefore, if it was also generated by a vectorial Lagrange multiplier, we would have zero degrees of freedom left!\footnote{Agreed: this argument rejects the possibility that the Lagrange multiplier itself be constrained; e.g.  a  divergence-free vector field. The main reason is that this would introduce a Lagrange multiplier for a Lagrange multiplier. Such symmetries are ``reducible'' and lie outside the scope of this investigation. If one wishes to completely obviate the need for this argument, one could simply  require that the gravitational field  have two propagating physical degrees of freedom; this already requires the remaining part of the Hamiltonian to be scalar.    } Thus, one further  assumption that goes into our general derivation of geometrodynamical evolution is that the theory is not topological.
 
Lastly, we require the Hamiltonian to depend at most quadratically on the momenta.
This can be justified by the need to avoid Ostrogradski instabilities (and thereby, an unphysical theory) that higher-than-quadratic momenta would engender in the corresponding Lagrangian theory (by inducing higher time derivatives). But for reasons of space, we will not give details; for a review, cf. \citet{Woodard_scholarpedia}. \\

These assumptions suffice to completely specify the Hamiltonian generating evolution. We now sketch the proof presented in  \citet{GomesShyam}.

\paragraph*{Statement of the results}
First,  we stipulate the same generator of spatial diffeomorphisms as in project (1), i.e. Hojman et al.'s \eqref{Diff_gen}: given the geometrodynamical phase space, this generator, $\mathscr{H}_\parallel$, is unique. 

The limited meshing relation we will require, namely, that the pure change part of the evolutions mesh in the right way, can be mathematically represented  as follows:
\be\label{eq:GS_geomequiv} \Delta^{12}_\perp\varphi=\Delta^{21}_\perp\varphi+\{\varphi, \mathscr{H}_\parallel\}(F(\xi_1, \xi_2)),\ee 
where $F$ is some fixed phase space function---about which we presuppose nothing---with the $\xi$ as arguments, and where we have identified:
\be \label{identifying}
\Delta^{12}_\perp\varphi:= \{\{ \varphi, \mathscr{H}_\perp(\xi^1_\perp)\}, \mathscr{H}_\perp(\xi^2_\perp)\}, \quad\text{and}\quad \Delta^{21}_\perp\varphi:= \{\{ \varphi, \mathscr{H}_\perp(\xi^2_\perp)\}, \mathscr{H}_\perp(\xi^1_\perp)\} \, .
\ee
Equation \eqref{eq:GS_geomequiv} guarantees that $\Delta^{12}_\perp\varphi\approx\Delta^{21}_\perp\varphi$ because we have taken $\mathscr{H}_\parallel$ to be the generator of spatial diffeomorphisms, and thus its action defines equivalence classes in  terms of spatial geometry.

Using the Jacobi identity, we find equation \eqref{eq:GS_geomequiv} holds if and only if 
 \be\label{eq:meshing_GS2}  \{\mathscr{H}_\perp(\xi^1_\perp), \mathscr{H}_\perp(\xi^2_\perp)\}=\mathscr{H}_\parallel (F(\xi_1, \xi_2))\ee
 for some phase space function $F$; (about which we again presuppose nothing, unlike Hojman et al, who demand that $\mathscr{H}_\parallel(F(\xi_1, \xi_2))$ reproduce the hypersurface deformation algebra). And moreover, if $F$ turns out to be phase space dependent, we will require a second condition, namely that  $\mathscr{H}_\parallel\approx 0$. \\

We first write a general scalar Hamiltonian generator as a sum of a kinetic and a potential term (both assumed to be non-identically zero):\footnote{In fact a bit more is assumed. If $V\equiv 0$, a slight generalization  of $T$ ensues in theorem \ref{theo:rigidity} (still no derivatives of the momenta, but more general contractions, i.e. an arbitrary DeWitt parameter); if $T\equiv 0$, any $V$ is trivially propagated. If $V\not\equiv 0$ but contains no derivatives of the metric (i.e. it is proportional to the scalar density $\sqrt{g}$), a linear term in the momenta in $T$ arises, but this change can be accommodated by the purely quadratic term through a canonical transformation which eliminates the potential term.} 
\be\label{eq:general}
\mathscr{H}_\perp[\pi^{ab}, h_{ab}; \sigma)=T[\pi^{ab}, h_{ab}; \sigma)+ V[ h_{ab}; \sigma) \; .
\ee
 Since the possible terms linear in the momentum can be generated by a canonical transformation of the variables, $T$ depends quadratically on the momenta.

The actual calculations are painstaking and cannot be meaningfully summarized here. For the purpose of illustration, we present the general form of the kinetic and potential terms, considered in \citet[p. 112503-3]{GomesShyam}.  The general kinetic term is of the form:
\begin{equation}
T[h,\pi;\sigma)=\sum_{r,m,n}\mathcal{B}_{abcd}^{i_{1}\cdots i_{n}\,j_{1}\cdots j_{m}}\left(\nabla_{i_{1}}\cdots\nabla_{i_{n}}\pi^{cd}\right)\left(\nabla_{j_{1}}\cdots\nabla_{j_{m}}\pi^{ab}\right)(\sigma) \; .
\label{eq:kinetic_original}
\end{equation}
The scalar function $V[h,x)$ and the tensor $\mathcal{B}_{abcd}^{i_{1}\cdots i_{n}\,j_{1}\cdots j_{m}}\mathcal{}$  depend on the metric $h_{ab}$
and its derivatives through $\ensuremath{R_{kl},\nabla_{a_{1}}R_{kl},\cdots,\nabla_{a_{1}}\cdots\nabla_{a_{r}}R_{kl}}$
where the number of explicit covariant derivatives, $r$, is arbitrary.
 Note that in all generality, the spatial derivative order of $\mathcal{B}_{abcd}^{i_{1}\cdots i_{n}\,j_{1}\cdots j_{m}}\mathcal{}$
and $V[h,\sigma)$ differ.

We can now state the theorem, proven in  \citet[p. 112503-23]{GomesShyam}:  \begin{theo}[Rigidity]\label{theo:rigidity}
Equation \eqref{eq:meshing_GS2} is satisfied by a unique generator in the family  \eqref{eq:general} (with T as defined in \eqref{eq:kinetic_original}). Namely:
\be\label{eqADMGS}
\mathcal{H}_{\text{\tiny{ADM}} \, \perp}[h,\pi;\sigma):=\left(a\epsilon\frac{\pi^{ab}\pi_{ab}-\frac{1}{2}\pi^2}{\sqrt{h}}-a^{-1}(R-2\Lambda)\sqrt{h}\right)(\sigma)\ee
where $a$ and $\Lambda$ are  arbitrary constants. 
\end{theo}

Thus, since  \eqref{eq:meshing_GS2} is obtained with an $F$ that is the same as given by the hypersurface deformation algebra of \eqref{vec_comm2} (and therefore $F$ is phase-space-dependent). Thus  \eqref{eq:GS_geomequiv} requires $ \mathscr{H}_\parallel\approx 0$. And since $\{ \mathscr{H}_\perp, \mathscr{H}_\parallel\}\propto \mathscr{H}_\perp$, dynamical preservation of $ \mathscr{H}_\parallel\approx 0$ implies also that $\mathscr{H}_\perp\approx 0$. \\

Thus we can now state:
\begin{cor}\label{cor:GS} Given an equivalence relation supplied by spatial diffeomorphisms, \eqref{Diff_gen}, and a generator of time evolution of the form \eqref{eq:general},  the only theory that enforces the meshing,\footnote{Of course, as in ADM, these definitions are unique only up to the relevant class of isomorphism of the theory in question: here, symplectomorphisms (or canonical transformations). } or path-independence condition, $\Delta^{12}_\perp\varphi\approx\Delta^{21}_\perp\varphi$, given in equation \eqref{eq:GS_geomequiv} is the ADM form of general relativity. \end{cor}

\section{A corollary: shape dynamics exonerated}\label{sec:payoff}
We turn to our positive corollary of Section \ref{sec:GS}. It is that another programme in foundations of classical gravity, shape dynamics, is innocent of an accusation that one might be tempted to make against it. The accusation arises from the fact that within a certain regime of shape dynamics, its Hamiltonian matches that of general relativity i.e. is the ADM Hamiltonian. This prompts the question, like I. Rabi’s quip when in 1936 the muon was discovered with its surprising properties: `who ordered that?’ That is: what is the motivation, from within the perspective of shape dynamics, for this form of Hamiltonian? Or more accusingly: is there none? Has shape dynamics `simply grabbed’ the ADM Hamiltonian from its illustrious rival, general relativity?

We will begin by recalling how the dialectic of such an accusation is often treated in philosophy: namely, in terms of a much-cited quip by Bertrand Russell condemning `theft’ and praising `honest toil’ (Section \ref{quip}). This review will not be just a philosophical reminiscence. Russell’s quip is unfair in a way that the `top theories’ $T_t$ in this paper’s three projects illustrate. Then we will turn to shape dynamics. We first introduce it (Section \ref{sd_expos}); and then  in Section \ref{sd_exon} we explain how, thanks to Section \ref{sec:GS}’s theorem, it is innocent of the accusation. It has indeed undertaken, not theft, but honest toil. 

\subsection{Russell’s quip}\label{quip}
Much of Russell’s writings in the period 1910 to 1930 proposed reductions in what Section \ref{funcredn} called the Nagelian sense: deduction of a `top theory’ $T_t$ from a `bottom theory’ $T_b$, augmented with extra premises we there labelled $B$ (for `bridge-laws’). Russell’s motivation, indeed inspiration, was of course the logicist programme of Frege, himself and Whitehead to try to reduce all of pure mathematics to logic: that is, to deduce pure mathematics from logic, augmented with appropriate definitions of mathematical words (e.g. `real number’) in terms of logic notions. 

In the light of that endeavour, especially his and Whitehead’s {\em magnum opus, Principia Mathematica}, Russell aimed to similarly reduce whole swathes of empirical knowledge, both everyday and technical, to theories about sense-experience. And in the logical empiricist school of the 1920s onwards, which was much influenced by logicism, other philosophers, such as Carnap, articulated similarly ambitious philosophical programmes to reduce the `problematic’ realm of `the theoretical’ to `observation’: programmes that led directly to fellow-empiricist Nagel’s formulation of reduction, as reviewed at the beginning of Section \ref{funcredn}. 

We need not linger on these philosophers’ aim of finding a firm foundation of empirical knowledge in sense-experience, or `pure observation’: an aim that was much criticized from about 1950 onwards.\footnote{\label{logicism}{Let alone endorse it: cf. footnote \ref{LewNotInstruml}. Nor need we linger on the shortcomings of logicism: in short, that the most it could claim to achieve is the deduction of pure mathematics from a single part of it, viz. set-theory, not from {\em logic}.}} What matters for us is these philosophers’ admitting the need to formulate judicious definitions that enable them to {\em deduce} the `problematic’ doctrine at issue. They are thus opposed to so-called {\em implicit definition}, i.e. to thinking it is enough, for justifying a concept or discourse, to give a set of postulates (`axioms') containing the concept and from which one's claims about it can be deduced. How do you know---they might say---that your deduction sets out from safe ground? 

Hence Russell's famous  quip about `theft over honest toil'. It is in his {\em Introduction to Mathematical Philosophy} (1919), in  his discussion of deducing the truths of arithmetic from logic, i.e. from set-theory (cf. footnote \ref{logicism}):
\begin{quote}
The method of `postulating' what we want has many advantages; they are the same as the advantages of theft over honest toil. Let us leave them to others and proceed with our honest toil. (1919, p. 71)
\end{quote}
Thus `theft' is here the dogmatic postulation of entities,  as being those things that obey certain axioms or postulates; and `toil' is here the work of finding judicious definitions so that the {\em definienda} can be shown  using logical inference alone to satisfy the claims made about them. 

Indeed, a {\em bon mot}; and much quoted. But it is unfair to dismiss making postulates at the level of the `top theory' as `theft'. In many cases, it is obviously hard work, even when one is given an agreed stock of words or concepts, to formulate axioms whose consequences express all that we wish to assert using those words or concepts. Think of the work involved in axiomatising Euclidean geometry: whether by Euclid and his contemporaries, or in our own era by Hilbert (in his {\em Grundlagen der Geometrie} of 1899) and his successors. Formulating an axiom system that implies all the desired theorems is an achievement; (especially if one requires, as Hilbert did, that the axioms be mutually independent). This is so, regardless of whether one has the philosopher’s or logician’s interest in giving underlying `bottom theory’ definitions of the objects in geometry’s subject-matter; e.g. of what a point in the plane {\em is} in terms of set-theory. 

We see this point also in the top theories $T_t$ of this paper’s three projects (and in our other spacetime examples in \citet{ButterfieldGomes_2}). For example, the formulation by Hojman et al. of their deformation algebra is an achievement, regardless of whether one goes on---as indeed they did---to give an underlying `bottom theory’ realization of it as a Hamiltonian constraint algebra. Similarly, the formulation by Schuller et al. of generalized geometries is an achievement, regardless of whether one goes on---as indeed they did---to give an underlying `bottom theory’ invoking predictivity and good behaviour conditions for matter and radiation, and showing a solution of the resulting consistency conditions.  Lastly, the formulation by Gomes and Shyam of path independence of evolution of spatial quantities is an achievement, regardless of whether one goes on to specify a Hamiltonian that has these properties.\footnote{Fortunately, the point is also noticed by philosophers of logic. \citet[p. 272]{OliverSmiley} call Russell's {\em bon mot} `one of the shoddiest slogans in philosophy’. And their reason is exactly this point: that writing down the right analysis of a concept, {\em before} any `reduction’ begins, often requires honest toil.  Thus they point out that  (i) Russell originally aimed it, unfairly, at Dedekind’s treatment of continuity, and (ii) `it assumes [wrongly] that we already know what we want … the examples from Dedekind show just how much honest toil it takes to discover—to formulate precisely---just what it is that we want’.}

So  much by way of general philosophy. We turn to shape dynamics: to urge that thanks to Section \ref{sec:GS}’s theorem, its use of the ADM Hamiltonian is {\em not} a case of theft. 

\subsection{A brief introduction to shape dynamics}\label{sd_expos}
This will be a very brief, and so incomplete, introduction to shape dynamics. For more technical details, see \citet[Ch. 12]{Flavio_tutorial} and \citet{SD_first, SD_linking}. 

\paragraph*{The relational motivation for shape dynamics}

Conceptually, shape dynamics is an offspring of relationism. Relationism starts from the idea  that all measurements are comparisons: that doubling the size
of all the rods and of the universe alongside them, will change nothing measurable; and  that speeding up all  clocks and physical
phenomena alongside  them, will change nothing measurable. No one has ever concretely measured (nor ever will measure) a
fundamentally dimensionful quantity.\footnote{See \citet{Julian_End} for an eloquent defence of these ideas.}

 But even if one accepts that  fundamentally  only relations are physically measured, general relativity associates a dimensionful quantity to trajectories (curves) in space-time, irrespectively of relations to any other object: namely, proper time (or distance). The \lq\lq{}experienced time\rq\rq{} along a trajectory in space-time depends only on the trajectory itself, without reference to anything else. (This  time is of course determined  only up to an affine transformation, reflecting a choice of unit and zero.)  So {\em prima facie}, general relativity fails to be relational, because of its physically significant but conformally {\em non}-invariant  \lq\lq{}experienced time\rq\rq{}  (i.e. proper time and distance).  Although  we believe that proper time should be a physical correlate of something relational, such as a ratio of numbers of oscillations of two atoms, the theory does not \textit{explicitly} require such an understanding. 
  
  So it is tempting to impose scale-invariance at the fundamental, theory-building level, as a criterion: to require theories to build all physical quantities from relations.  
Over the years, prompted by this sort of relational rationale, many consistent theories of gravity with 4--dimensional space-time scale-invariance have been proposed.\footnote{Such models come under different names (e.g. conformal gravity and  Weyl gravity). Their appeal lies in their renormalization properties, see e.g. \citet{Stelle_grav, Fradkin_conf}.} But, so far as we know, none has overcome this shortcoming of general relativity, while keeping its explanatory power. 

  But if we accept the ontological `3+1' split of spacetime into space and time, we can perhaps break this {\em impasse}. For we now gain access to a different kind of symmetry:  those that act on each spatial field configuration. Spatial diffeomorphisms  enforce the notion that location is relational; i.e. positions only make sense relative to other positions. Analogously, 
spatial conformal invariance is  a relational symmetry: it means that local scale can only be made sense of relationally, i.e. by comparing one size to another. 

\paragraph*{The technical argument for shape dynamics.}

In brief, the technical idea behind shape dynamics, is that the ADM formulation of general relativity, with its refoliation redundancy (represented by $\mathscr{H}_{\text{\tiny ADM}\perp}\approx 0$), is dual to a theory that possesses spatial conformal invariance. The reason such a dual formulation is possible, is that the constraint that is the symplectic generator of conformal transformations, is first class with respect to diffeomorphisms, and can be seen as a gauge-fixing of the symmetries generated by $\mathscr{H}_{\text{\tiny ADM}\perp}$. Thus we can represent the ADM Hamiltonian in an alternative (non-local) form in the space of conformal geometries. The details of this duality are spelled out in \citet{SD_linking}. Here, for brevity, we will only describe how the theory is closely related to what is known as the `York method' for solving general relativity's initial value constraints. 

We start from the fact that in the ADM formulation, not all initial data $(h_{ab}, \pi_{ab})$ are acceptable.  To find initial data, we must first solve the two ADM constraints, $\mathscr{H}_\perp=\mathscr{H}_\parallel=0$. 
  The York method \citet{York, York3} is the only known mathematical tool for generically  attacking the initial value problem of general relativity \citet{3+1_book}. Indeed,  most of the formal proofs of existence and uniqueness of solutions of general relativistic {dynamics} require the use of a particular choice of foliation, namely that the mean (of the trace of the) extrinsic curvature (CMC) be constant throughout the given leaf. 
   And as we shall see in a moment, this foliation is the one that is compatible with the existence of a shape dynamics solution. 
As a matter of fact, every test that has ever been passed by GR is known to be consistent with a CMC foliation.\footnote{ The foliation does break down inside known solutions with event horizons. But, in the generic case where event horizons form, Marsden and Tipler conjecture that 
\lq\lq{}the [CMC] foliation does fill out at least the exterior of the event horizon of the singularity, and that the foliation fills out the  entire spacetime if it is geodesically complete \rq\rq{}\citet[p. 126]{Tipler_CMC}.
  }
   
  More importantly for us, it turns out that this choice  has special properties: it allows  a  description of the  evolution of spatial geometry  without reference to local spatial scale. That is, it allows one to describe gravitational evolution in the auxiliary time $\tau$, \footnote{Which, from \eqref{eq:CMC_ctraint}, is just the average of the trace of the momenta, $\tau:=\frac{\int  \text{tr} \pi}{V(h)}$, where $V(h):=\int \sqrt h$.} in terms of spatially \lq{}conformally invariant\rq{} geometries.      This is because the constraints that these foliations are obliged to satisfy in order to be CMC, namely, 
 \be\label{eq:CMC_ctraint}\mathcal{C}=  \text{tr} \, \pi-\tau\sqrt{h}=0,\ee
 where $ \text{tr} \, \pi=h_{ab}\pi^{ab} $ and $\tau$ is a constant, generate,  via Noether\rq{}s theorem,  canonical transformations.  Such transformations are exactly changes of spatial scale, i.e.  the local conformal transformations.\footnote{In fact, the momenta are set to be constant trace, not trace-free; which would be the true generator of conformal transformations.  However, one can interpret the constant trace part as defining an auxiliary quantity, the York time. The transverse traceless choice sets the gravitational momenta to be purely tensorial (spin-2), with no scalar (spin-0) or vector (spin-1) components.}  These results show that surprisingly,  although general relativity is not fundamentally concerned with spatial conformal  geometries, it is deeply related to them.
  
\begin{figure}[h!]\label{fig:ctraint}
\center
\includegraphics[width=0.4\textwidth]{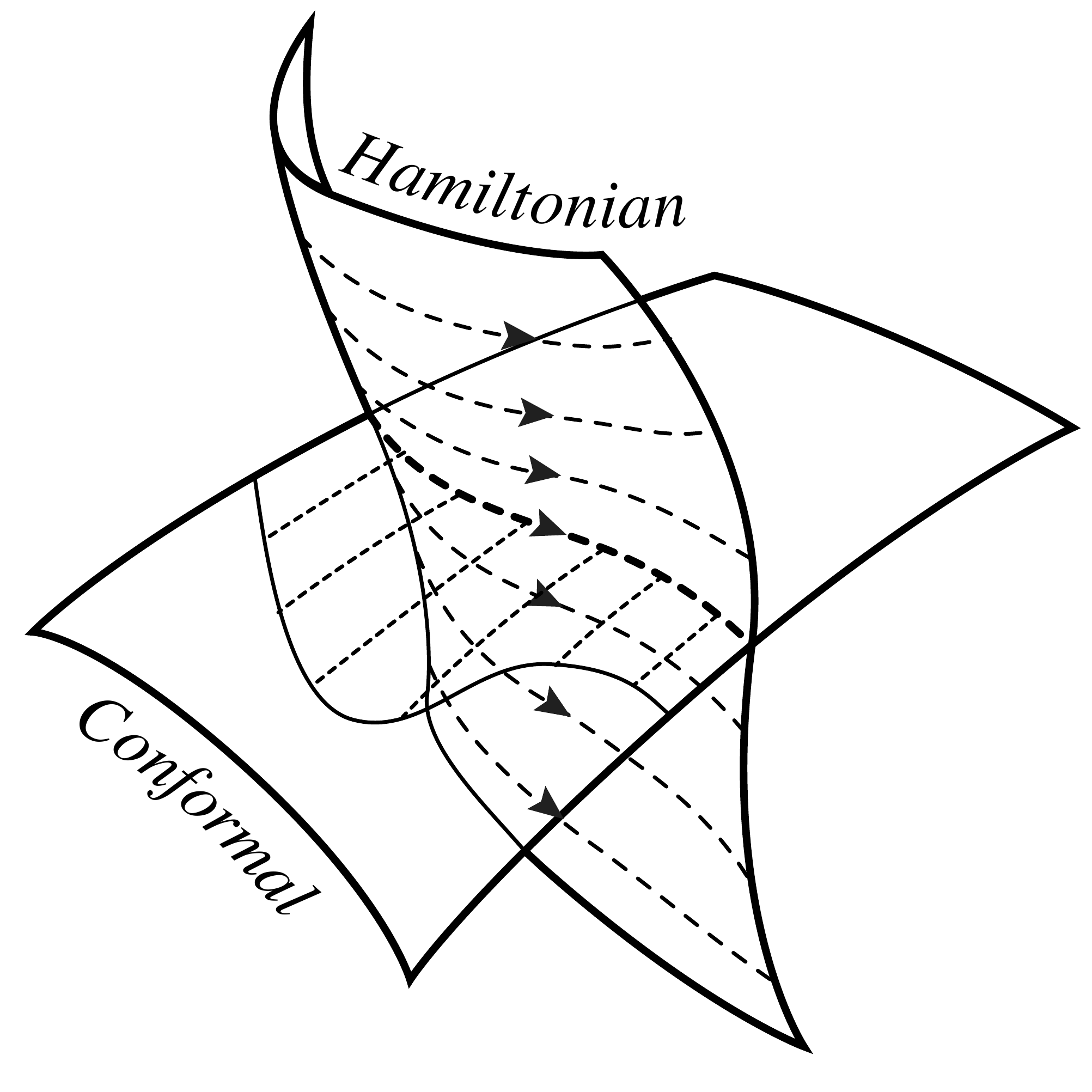}
\caption{A schematic representation of the phase space of general relativity. In it, two
constraints coexist, which are good gauge-fixings for each other and are both
first-class with respect to the diffeomorphism constraint. One is the Hamiltonian
constraint and the other is the conformal (Weyl) constraint. The Hamiltonian
constraint is completely gauge-fixed by the conformal constraint except
for a single residual global constraint. It Poisson-commutes with the conformal
constraint and generates a vector 
flow on the Hamiltonian constraint surface
(represented in the figure), which is parallel to the conformal constraint surface.
This vector 
flow generates the time evolution of the system in the intersection
between the two surfaces. Any solution can then be represented in an arbitrary
conformal gauge by lifting it from the intersection to an arbitrary curve on
the conformal constraint surface. All such lifted curves are gauge-equivalent
solutions of a conformal gauge theory with conformally-invariant Hamiltonian.}
\end{figure}

  Schematically, the trade can be represented, in the jargon of Dirac constraint analysis,  as two first class constraint surfaces  being fully second-class with respect to each other. Being `first class' means that constraints weakly commute, that is, that they close on-shell and thus can be taken as symmetry generating; `second class' means that they do not weakly commute, indeed, their Poisson bracket forms an invertible matrix, and thus one phase space function serves as a gauge-fixing for the other. See figure \ref{fig:ctraint}.
  
  These properties mean that we can split $\mathscr{H}_\perp$ into a local and a global part. The local constraint generates a symmetry that can be traded with local conformal invariance, whereas the global part generates a non-local Hamiltonian in conformal superspace, that we can write, in the unconstrained phase space as: 
\be \mathscr{H}_{\text{\tiny global}}=\int (\Omega^6(h, \pi)-1)\sqrt{h}\ee
where $\Omega$ is the York conformal factor (described in appendix \ref{app:York}).

Thus $\mathscr{H}_\perp-\mathscr{H}_{\text{\tiny global}}$ is entirely second class with respect to $\mathcal{C}$, and we have two separate first class systems of local constraints: the standard ADM one, given by  $\mathscr{H}_\perp\approx 0$ and $\mathscr{H}_\parallel\approx0$,  and the shape dynamics one, 
\be \mathcal{C}\approx 0, \qquad \mathscr{H}_\parallel\approx0 \; : \ee 
which are  both conserved by the Hamiltonian $ \mathscr{H}_{\text{\tiny global}}$. Moreover, the dynamics of this system  matches  that of ADM general relativity (whenever a CMC foliation is available, as discussed above).

In sum, shape dynamics amounts to taking seriously these hints: hints from the initial value formulation of general relativity and from considering symmetries compatible with  instantaneous physical configuration states. 
The result is a theory about the dynamics of space that is based on different symmetry principles than the standard diffeomorphism or Lorentz  invariance of spacetime. Namely, the symmetries of the theory are taken to be conformal diffeomorphisms. Indeed, starting from the broad idea that shape dynamics is to be a theory about space, it is natural to focus on the space Riem, of unconstrained spatial 3-metrics on a 3-manifold, $\Sigma$. Then one can  justify the symmetry group of spatial conformal diffeomorphisms as being  the maximal local group of symmetries acting on Riem \citet{Gomes2017}. Thus we get the idea of conformal superspace, which has been shown to have a well-defined symplectic reduction \citet{FiMa77}. 

But as we have also seen, it has one peculiar feature:  a non-local generator of evolution (Hamiltonian). And if we focus on conformal superspace, there are admittedly many Hamiltonians on it that look more natural  than the non-local one that shape dynamics officially adopts: viz. the Hamiltonian obtained by extending general relativity's ADM Hamiltonian to this space. Agreed: this official choice enables shape dynamics to match (in many circumstances) the successes of general relativity.\footnote{ Fortunately, we can in many circumstances find explicit solutions of shape dynamics, albeit at the cost of simplifying the problem by assuming it is highly symmetric: see \citet[Ch. 13]{Flavio_tutorial}, and references therein.} But the choice looking unnatural triggers the accusation of `theft' ... 

\subsection{Shape dynamics exonerated}\label{sd_exon}

But Section \ref{sec:GS}'s  Theorem \ref{theo:rigidity}  exonerates shape dynamics. For from most such natural Hamiltonians on conformal superspace, we would {\em not}  be able to deduce appropriate relations to `local change', or duration in the sense of Section \ref{constrduratn}. 

To put the point more specifically and more positively, in terms of functionalist reduction: the shape dynamics Hamiltonian is the nearest realizer on (the cotangent bundle of) conformal superspace of a Hamiltonian possessing a representative that---that is: that in some gauge---yields  the notion of duration which we formulated in Section \ref{constrduratn}'s $T_t$. That is: this Hamiltonian gives, in some nearest realization, a reduction, in the vocabulary of conformal superspace ($T_b$), of the $T_t$ notion of time. 

That, through Theorem \ref{theo:rigidity}, this can be done without invoking spacetime ideas---in particular, without invoking the relativity of simultaneity, or its more sophisticated cousin, refoliation invariance---is, we submit, remarkable. And it is essential for the basic outlook of shape dynamics, i.e. an ontology based on conformal superspace, to be tenable.

 In sum: shape dynamics is exonerated as the unique theory in conformal superspace that can be lifted to match the dynamics of a theory of spatial metrics with a local notion of duration.

\section{Conclusions}\label{sec:concl}

We shall not here summarise the three geometrodynamical  projects, (1) to (3), that we have reviewed and connected to functionalist reduction. For  Section \ref{appetizer}  summarised them in themselves, and Section \ref{funcredn} summarised them as illustrations of functionalist reduction. Let us instead give a table, allowing one to compare at a glance these three projects, and also shape dynamics.

\bigskip

\begin{tabular}{ |p{3cm}||p{3cm}|p{3cm}|p{3cm}|p{3cm}|  }
 \hline
 \multicolumn{5}{|c|}{The projects:  their assumptions, results and uniqueness} \\
 \hline
 &  spacetime manifold? & Spacetime metric?& Recover Einsteinian dyn.? & Uniqueness theorem?  \\
 \hline
Hojman et al.  & $\checkmark$    &$\checkmark$&   $\checkmark$ & $\checkmark$ \\
Schuller et al.  &   $\checkmark$  & $\times$   &$?$ &$\times$\\
  Gomes-Shyam   &$\times$ & $\times$& $\checkmark$ & $\checkmark$\\
   Shape dynam.   &$\times$ & $\times$& $\checkmark$ & $\checkmark^{*}$\\
 \hline
\end{tabular}\\
 
\bigskip

This table, with its brief labels, summarizes in the first two columns the {\em assumptions}, and in the last two columns the {\em results}, of functional reduction. It is of course a simplification of the situation. For example: although project (2) (by Schuller et al.) does not assume a Lorentzian metric, it does use some (dynamical) notion of a coupling metric, or geometry (the tensor $G$). On the other hand, the purely spatial ontology of project (3) \citet{GomesShyam} has a spatial Riemannian metric in its fundamental variables.  We have also put an asterisk in the last item (uniqueness theorem for shape dynamics), because (as discussed in Section \ref{sd_exon}) it is the unique theory in conformal superspace that can be lifted to match the dynamics of a theory of spatial metrics with a local notion of duration. That is: the asterisk signals the fact that it does not have a uniqueness theorem solely in terms of conformal superspace. The question mark in the third column, for project (2), is correlated to a lack of uniqueness: for we do not know if there are matter fields which would allow a different sort of spacetime (and spacetime dynamics) through the constructions of that project.\\

In this paper, we have used the philosophical idea of functionalist reduction to characterize time in three broadly geometrodynamical---but purely classical---theories. We thus arrive, in effect, at the threshold of one of H-D. Zeh’s main endeavours: to understand time in quantum geometrodynamics. That is of course a vast endeavour, in which Zeh, his many collaborators such as Kiefer \citet{Kiefer_book}, and of course many others, have done a great deal of work.
   We obviously cannot go into this here: (for  masterly introductions to the programme, cf. \citet[Ch. 6]{Zeh_time}  and 
\citet[Ch. 5]{Kiefer_book}). In closing, let us just make a hopeful remark on the possible use of our philosophical notions of functional role, and functionalist reduction.

We recall that in this endeavour, one main aim is to extract a time variable, with respect to which the matter degrees of freedom would obey a time-dependent Schr\"{o}dinger equation, from a WKB-like  approximation to the (fundamentally timeless) Wheeler-DeWitt equation. We can think of this as an internal time being identified by its functional role: the matter degrees of freedom are functionally differentiated, and  identifying the functional coefficient with time yields a Schrodinger-like equation for these degrees of freedom. Accordingly, our discussion suggests that one might explore defining  time  by other possible functional roles. But of course, whether any such definition will prove useful in the context of quantum geometrodynamics remains to be seen. We can only hope that the functionalist reductive perspective we have offered here for the classical case may prove useful in people’s effort to better understand the quantum case ... long may Hans-Dieter Zeh’s scientific creativity and craftsmanship inspire that effort!\\

\bigskip

\subsection*{Acknowledgements}
 We thank: (i) audiences in seminars at the Black Hole Initiative, Harvard, LSE, MIT and Muenchen, for comments; and (ii) Laurenz Hudetz, Nick Huggett, Tushar Menon and Brian Pitts, for correspondence.

\begin{appendix}
\section{The York method}\label{app:York}
Let us briefly review the York  method, by which general relativity's constraints  decouple and turn into elliptic equations. On a closed spatial hypersurface, we start with the (by-now standard!)  metric $h_{ab}$ and tensor density $\pi^{ab}$; but also with  a real number ${\tau}$.  The York method then consists in the following steps:  
  
\begin{flushleft}
1. Given a traceless tensor, one can  find its projection to a transverse component. That is, upon the substitution $\pi^{ab}\rightarrow \pi^{ab}-\nabla^{(a} {\chi^{j)}} $ (where round brackets signify index symmetrization, $A_{(ab)} := A_{ab}+A_{ba}$) and $\nabla$ is the Levi-Civita covariant derivative with respect to $h_{ab}$,  one can solve the transversality condition (with respect to the vector field ${\chi^a}$):
$$
\nabla_b (\nabla^{(a} {\chi^{b)}} - {\frac 2 3} h^{ab} \nabla_c{\chi^c}) = \nabla_b(\pi^{ab} -{\frac 1 3} \text{tr} \, \pi ~ h^{ab}) ~ .
$$

Thereby we obtain the transverse-traceless momentum ${\pi_{\mbox{\tiny TT}}^{ab}} $:
$$
{\pi^{ab}_{\mbox{\tiny TT}}} = (\pi^{ab} -{\frac 1 3} h^{ab} \text{tr} \, \pi) - (\nabla^a {\chi^b} +\nabla^b {\chi^a} - {\frac 2 3} h^{ab} \nabla_c {\chi^c})\, \; .
$$
This means that for any constant $\tau$, the momentum $p^{ab}:={\pi^{ab}_{\mbox{\tiny TT}}}+{\frac 1 2} {\tau} \, h^{ab} \, \sqrt{h} $ satisfies $\mathscr{H}_\parallel=0$. Importantly, it represents a \emph{CMC slicing} (constant mean curvature), since its trace is necessarily constant:
$$
 h_{ab} p^{ab} = {{\frac 3 2} \tau} \sqrt{h} = \text{\it const.}~\sqrt{h} ~ .
$$

2.  Upon \textit{conformally transforming}  the canonical variables, as $(h_{ab}, {\pi^{ab}_{\mbox{\tiny TT}}})\rightarrow (\Omega^{4} h_{ab}, \Omega^{-4}{\pi^{ab}_{\mbox{\tiny TT}}})$, from the scalar constraint $\mathscr{H}_\perp=0$ one obtains the Lichnerowicz--York equation (LY) with respect to the scalar field ${\Omega}$:
$$
 \frac {{\Omega^{-6}}} {\sqrt h}{\pi^{ab}_{\mbox{\tiny TT}}}^2-{\frac 3 8} \sqrt h \, {\Omega^6} {\tau^2} - \sqrt h \left( R \,{\Omega^2} - 8 \, {\Omega} \Delta {\Omega} \right)  = 0~.
$$
This is an elliptic equation which can, under certain additional assumptions,  be solved for $\Omega$: thereby satisfying all of the ADM constraints on a CMC slice $\left( \text{tr} \, {p}  = {\frac 3 2} \, {\tau}  \,{ \sqrt{{h}}}\right)$. The metric and momenta satisfying the initial value problem are then given by $p^{ab}$ and $\gamma_{ab}=\Omega^{4} h_{ab}$. 
\end{flushleft}

One important point hidden in the analysis is that  the LY equation and the transversality condition
are conformally invariant:
$$
\left\{\begin{array}{l}
h_{ab} \to e^{4\phi}h_{ab}\\
\pi^{ab} \to e^{-4\phi}\pi^{ab}
\end{array}\right.
 ~~~~ \Rightarrow ~~~~
 \left\{\begin{array}{l}
\gamma_{ab}[e^{4\phi}h_{ab}, e^{-4\phi}\pi^{ab}] = \gamma_{ab}[h, \pi]\\
p^{ab}[e^{4\phi}h_{ab}, e^{-4\phi}\pi^{ab}] = p^{ab}[h_{ab}, \pi^{ab}]  \; .
\end{array}\right.
$$

\end{appendix}


\begin{thebibliography}{}

\bibitem [\protect \citeauthoryear {%
Arnowitt%
, Deser%
\BCBL {}\ \BBA {} Misner%
}{%
Arnowitt%
\ \protect \BOthers {.}}{%
{\protect \APACyear {1962}}%
}]{%
ADM}
\APACinsertmetastar {%
ADM}%
\begin{APACrefauthors}%
Arnowitt, R.%
, Deser, S.%
\BCBL {}\ \BBA {} Misner, C.%
\end{APACrefauthors}%
\unskip\
\newblock
\APACrefYearMonthDay{1962}{}{}.
\newblock
{\BBOQ}\APACrefatitle {{The Dynamics of General Relativity, pp. 227-264}} {{The
  Dynamics of General Relativity, pp. 227-264}}.{\BBCQ}
\newblock
\BIn{} \APACrefbtitle {{Gravitation: an introduction to current research, L.
  Witten, ed.}} {{Gravitation: an introduction to current research, L. Witten,
  ed.}}
\newblock
\APACaddressPublisher{}{Wiley, New York}.
\PrintBackRefs{\CurrentBib}

\bibitem [\protect \citeauthoryear {%
Barbour%
}{%
Barbour%
}{%
{\protect \APACyear {1999}}%
}]{%
Julian_End}
\APACinsertmetastar {%
Julian_End}%
\begin{APACrefauthors}%
Barbour, J.%
\end{APACrefauthors}%
\unskip\
\newblock
\APACrefYear{1999}.
\newblock
\APACrefbtitle {{The End of Time: The Next Revolution in Physics}} {{The End of
  Time: The Next Revolution in Physics}}.
\newblock
\APACaddressPublisher{}{Oxford University Press}.
\PrintBackRefs{\CurrentBib}

\bibitem [\protect \citeauthoryear {%
Barbour%
\ \BBA {} Bertotti%
}{%
Barbour%
\ \BBA {} Bertotti%
}{%
{\protect \APACyear {1982}}%
}]{%
Barbour_Bertotti}
\APACinsertmetastar {%
Barbour_Bertotti}%
\begin{APACrefauthors}%
Barbour, J.%
\BCBT {}\ \BBA {} Bertotti, B.%
\end{APACrefauthors}%
\unskip\
\newblock
\APACrefYearMonthDay{1982}{}{}.
\newblock
{\BBOQ}\APACrefatitle {{Mach's Principle and the Structure of Dynamical
  Theories}} {{Mach's Principle and the Structure of Dynamical
  Theories}}.{\BBCQ}
\newblock
\APACjournalVolNumPages{Proceedings of the Royal Society of London A:
  Mathematical, Physical and Engineering Sciences}{382}{1783}{295--306}.
\newblock
\begin{APACrefURL}
  \url{http://rspa.royalsocietypublishing.org/content/382/1783/295}
  \end{APACrefURL}
\newblock
\begin{APACrefDOI} \doi{10.1098/rspa.1982.0102} \end{APACrefDOI}
\PrintBackRefs{\CurrentBib}

\bibitem [\protect \citeauthoryear {%
Barbour%
, Foster%
\BCBL {}\ \BBA {} O'Murchadha%
}{%
Barbour%
\ \protect \BOthers {.}}{%
{\protect \APACyear {2002}}%
}]{%
Barbour_RWR}
\APACinsertmetastar {%
Barbour_RWR}%
\begin{APACrefauthors}%
Barbour, J.%
, Foster, B\BPBI Z.%
\BCBL {}\ \BBA {} O'Murchadha, N.%
\end{APACrefauthors}%
\unskip\
\newblock
\APACrefYearMonthDay{2002}{}{}.
\newblock
{\BBOQ}\APACrefatitle {{Relativity without relativity}} {{Relativity without
  relativity}}.{\BBCQ}
\newblock
\APACjournalVolNumPages{Class. Quant. Grav.}{19}{}{3217-3248}.
\newblock
\begin{APACrefDOI} \doi{10.1088/0264-9381/19/12/308} \end{APACrefDOI}
\PrintBackRefs{\CurrentBib}

\bibitem [\protect \citeauthoryear {%
Braddon-Mitchell%
\ \BBA {} Nola%
}{%
Braddon-Mitchell%
\ \BBA {} Nola%
}{%
{\protect \APACyear {2009}}%
}]{%
MitchellNola}
\APACinsertmetastar {%
MitchellNola}%
\begin{APACrefauthors}%
Braddon-Mitchell, D.%
\BCBT {}\ \BBA {} Nola, R.%
\end{APACrefauthors}%
\unskip\
\newblock
\APACrefYear{2009}.
\newblock
\APACrefbtitle {{Conceptual Analysis and Philosophical Naturalism}}
  {{Conceptual Analysis and Philosophical Naturalism}}.
\newblock
\APACaddressPublisher{}{MIT Press: Bradford Books.}
\PrintBackRefs{\CurrentBib}

\bibitem [\protect \citeauthoryear {%
Brown%
}{%
Brown%
}{%
{\protect \APACyear {2006}}%
}]{%
Brown_book}
\APACinsertmetastar {%
Brown_book}%
\begin{APACrefauthors}%
Brown, H.%
\end{APACrefauthors}%
\unskip\
\newblock
\APACrefYear{2006}.
\newblock
\APACrefbtitle {{Physical Relativity}} {{Physical Relativity}}.
\newblock
\APACaddressPublisher{}{Oxford University Press.}
\PrintBackRefs{\CurrentBib}

\bibitem [\protect \citeauthoryear {%
Butterfield%
}{%
Butterfield%
}{%
{\protect \APACyear {1989}}%
}]{%
Butterfield_hole}
\APACinsertmetastar {%
Butterfield_hole}%
\begin{APACrefauthors}%
Butterfield, J.%
\end{APACrefauthors}%
\unskip\
\newblock
\APACrefYearMonthDay{1989}{}{}.
\newblock
{\BBOQ}\APACrefatitle {{The Hole Truth}} {{The Hole Truth}}.{\BBCQ}
\newblock
\APACjournalVolNumPages{The British Journal for the Philosophy of
  Science}{40}{1}{1-28}.
\newblock
\begin{APACrefURL} \url{https://doi.org/10.1093/bjps/40.1.1} \end{APACrefURL}
\newblock
\begin{APACrefDOI} \doi{10.1093/bjps/40.1.1} \end{APACrefDOI}
\PrintBackRefs{\CurrentBib}

\bibitem [\protect \citeauthoryear {%
Butterfield%
}{%
Butterfield%
}{%
{\protect \APACyear {1992}}%
}]{%
Butterfield92}
\APACinsertmetastar {%
Butterfield92}%
\begin{APACrefauthors}%
Butterfield, J.%
\end{APACrefauthors}%
\unskip\
\newblock
\APACrefYearMonthDay{1992}{}{}.
\newblock
{\BBOQ}\APACrefatitle {{David Lewis Meets John Bell}} {{David Lewis Meets John
  Bell}}.{\BBCQ}
\newblock
\APACjournalVolNumPages{Philosophy of Science, {\bf 59}, 26-43.}{}{}{}.
\newblock
\begin{APACrefDOI} \doi{10.1086/289652} \end{APACrefDOI}
\PrintBackRefs{\CurrentBib}

\bibitem [\protect \citeauthoryear {%
Butterfield%
}{%
Butterfield%
}{%
{\protect \APACyear {2004}}%
}]{%
Butterfield_DLHJ}
\APACinsertmetastar {%
Butterfield_DLHJ}%
\begin{APACrefauthors}%
Butterfield, J.%
\end{APACrefauthors}%
\unskip\
\newblock
\APACrefYearMonthDay{2004}{}{}.
\newblock
{\BBOQ}\APACrefatitle {{David Lewis Meets Hamilton and Jacobi}} {{David Lewis
  Meets Hamilton and Jacobi}}.{\BBCQ}
\newblock
\APACjournalVolNumPages{Philosophy of Science}{71}{5}{1095--1106}.
\newblock
\begin{APACrefURL} \url{https://www.jstor.org/stable/10.1086/428013}
  \end{APACrefURL}
\PrintBackRefs{\CurrentBib}

\bibitem [\protect \citeauthoryear {%
Butterfield%
}{%
Butterfield%
}{%
{\protect \APACyear {2011}}%
}]{%
Butterfield_emergence}
\APACinsertmetastar {%
Butterfield_emergence}%
\begin{APACrefauthors}%
Butterfield, J.%
\end{APACrefauthors}%
\unskip\
\newblock
\APACrefYearMonthDay{2011}{}{}.
\newblock
{\BBOQ}\APACrefatitle {Emergence, Reduction and Supervenience: A Varied
  Landscape} {Emergence, reduction and supervenience: A varied
  landscape}.{\BBCQ}
\newblock
\APACjournalVolNumPages{Foundations of Physics {\bf 41}, 920–959.}{}{}{}.
\PrintBackRefs{\CurrentBib}

\bibitem [\protect \citeauthoryear {%
Butterfield%
\ \BBA {} Gomes%
}{%
Butterfield%
\ \BBA {} Gomes%
}{%
{\protect \APACyear {2020}}%
{\protect \APACexlab {{\protect \BCnt {1}}}}}]{%
ButterfieldGomes_1}
\APACinsertmetastar {%
ButterfieldGomes_1}%
\begin{APACrefauthors}%
Butterfield, J.%
\BCBT {}\ \BBA {} Gomes, H.%
\end{APACrefauthors}%
\unskip\
\newblock
\APACrefYearMonthDay{2020{\protect \BCnt {1}}}{}{}.
\newblock
{\BBOQ}\APACrefatitle {Functionalism as a species of reduction} {Functionalism
  as a species of reduction}.{\BBCQ}
\newblock
\BIn{} \APACrefbtitle {{Current Debates in Philosophy of Science: In Honor of
  Roberto Torretti}.} {{Current Debates in Philosophy of Science: In Honor of
  Roberto Torretti}.}
\newblock
\APACaddressPublisher{}{Springer}.
\PrintBackRefs{\CurrentBib}

\bibitem [\protect \citeauthoryear {%
Butterfield%
\ \BBA {} Gomes%
}{%
Butterfield%
\ \BBA {} Gomes%
}{%
{\protect \APACyear {2020}}%
{\protect \APACexlab {{\protect \BCnt {2}}}}}]{%
ButterfieldGomes_2}
\APACinsertmetastar {%
ButterfieldGomes_2}%
\begin{APACrefauthors}%
Butterfield, J.%
\BCBT {}\ \BBA {} Gomes, H.%
\end{APACrefauthors}%
\unskip\
\newblock
\APACrefYearMonthDay{2020{\protect \BCnt {2}}}{}{}.
\newblock
{\BBOQ}\APACrefatitle {Spacetime functionalism {\em avant la lettre}}
  {Spacetime functionalism {\em avant la lettre}}.{\BBCQ}
\newblock
\APACjournalVolNumPages{Forthcoming}{}{}{}.
\PrintBackRefs{\CurrentBib}

\bibitem [\protect \citeauthoryear {%
Button%
\ \BBA {} Walsh%
}{%
Button%
\ \BBA {} Walsh%
}{%
{\protect \APACyear {2018}}%
}]{%
ButtonWalsh}
\APACinsertmetastar {%
ButtonWalsh}%
\begin{APACrefauthors}%
Button, T.%
\BCBT {}\ \BBA {} Walsh, S.%
\end{APACrefauthors}%
\unskip\
\newblock
\APACrefYear{2018}.
\newblock
\APACrefbtitle {{Philosophy and Model Theory}} {{Philosophy and Model Theory}}.
\newblock
\APACaddressPublisher{}{Oxford University Press}.
\PrintBackRefs{\CurrentBib}

\bibitem [\protect \citeauthoryear {%
Callender%
}{%
Callender%
}{%
{\protect \APACyear {2017}}%
}]{%
Callender_book}
\APACinsertmetastar {%
Callender_book}%
\begin{APACrefauthors}%
Callender, C.%
\end{APACrefauthors}%
\unskip\
\newblock
\APACrefYear{2017}.
\newblock
\APACrefbtitle {{What Makes Time Special}} {{What Makes Time Special}}.
\newblock
\APACaddressPublisher{}{Oxford University Press}.
\PrintBackRefs{\CurrentBib}

\bibitem [\protect \citeauthoryear {%
Dizadji-Bahmani%
, Frigg%
\BCBL {}\ \BBA {} Hartmann%
}{%
Dizadji-Bahmani%
\ \protect \BOthers {.}}{%
{\protect \APACyear {2010}}%
}]{%
Frigg_Nagelian}
\APACinsertmetastar {%
Frigg_Nagelian}%
\begin{APACrefauthors}%
Dizadji-Bahmani, F.%
, Frigg, R.%
\BCBL {}\ \BBA {} Hartmann, S.%
\end{APACrefauthors}%
\unskip\
\newblock
\APACrefYearMonthDay{2010}{}{}.
\newblock
{\BBOQ}\APACrefatitle {{Who's afraid of Nagelian reduction?}} {{Who's afraid of
  Nagelian reduction?}}{\BBCQ}
\newblock
\APACjournalVolNumPages{Erkenntnis, {\bf 73}, 393-412.}{}{}{}.
\PrintBackRefs{\CurrentBib}

\bibitem [\protect \citeauthoryear {%
Düll%
, Schuller%
, Stritzelberger%
\BCBL {}\ \BBA {} Wolz%
}{%
Düll%
\ \protect \BOthers {.}}{%
{\protect \APACyear {2018}}%
}]{%
Dull}
\APACinsertmetastar {%
Dull}%
\begin{APACrefauthors}%
Düll, M.%
, Schuller, F\BPBI P.%
, Stritzelberger, N.%
\BCBL {}\ \BBA {} Wolz, F.%
\end{APACrefauthors}%
\unskip\
\newblock
\APACrefYearMonthDay{2018}{}{}.
\newblock
{\BBOQ}\APACrefatitle {{Gravitational closure of matter field equations}}
  {{Gravitational closure of matter field equations}}.{\BBCQ}
\newblock
\APACjournalVolNumPages{Phys. Rev.}{D97}{8}{084036}.
\newblock
\begin{APACrefDOI} \doi{10.1103/PhysRevD.97.084036} \end{APACrefDOI}
\PrintBackRefs{\CurrentBib}

\bibitem [\protect \citeauthoryear {%
Fischer%
\ \BBA {} Marsden%
}{%
Fischer%
\ \BBA {} Marsden%
}{%
{\protect \APACyear {(1977)}}%
}]{%
FiMa77}
\APACinsertmetastar {%
FiMa77}%
\begin{APACrefauthors}%
Fischer, A.%
\BCBT {}\ \BBA {} Marsden, J.%
\end{APACrefauthors}%
\unskip\
\newblock
\APACrefYearMonthDay{(1977)}{}{}.
\newblock
{\BBOQ}\APACrefatitle {The manifold of conformally equivalent metrics} {The
  manifold of conformally equivalent metrics}.{\BBCQ}
\newblock
\APACjournalVolNumPages{Can. J. Math.}{29}{}{193-209}.
\PrintBackRefs{\CurrentBib}

\bibitem [\protect \citeauthoryear {%
Fradkin%
\ \BBA {} Tseytlin%
}{%
Fradkin%
\ \BBA {} Tseytlin%
}{%
{\protect \APACyear {1985}}%
}]{%
Fradkin_conf}
\APACinsertmetastar {%
Fradkin_conf}%
\begin{APACrefauthors}%
Fradkin, E.%
\BCBT {}\ \BBA {} Tseytlin, A.%
\end{APACrefauthors}%
\unskip\
\newblock
\APACrefYearMonthDay{1985}{}{}.
\newblock
{\BBOQ}\APACrefatitle {Conformal supergravity} {Conformal supergravity}.{\BBCQ}
\newblock
\APACjournalVolNumPages{Physics Reports}{119}{4}{233 - 362}.
\newblock
\begin{APACrefURL}
  \url{http://www.sciencedirect.com/science/article/pii/0370157385901383}
  \end{APACrefURL}
\newblock
\begin{APACrefDOI} \doi{https://doi.org/10.1016/0370-1573(85)90138-3}
  \end{APACrefDOI}
\PrintBackRefs{\CurrentBib}

\bibitem [\protect \citeauthoryear {%
Giesel%
, Schuller%
, Witte%
\BCBL {}\ \BBA {} Wohlfarth%
}{%
Giesel%
\ \protect \BOthers {.}}{%
{\protect \APACyear {2012}}%
}]{%
Giesel_Schuller}
\APACinsertmetastar {%
Giesel_Schuller}%
\begin{APACrefauthors}%
Giesel, K.%
, Schuller, F\BPBI P.%
, Witte, C.%
\BCBL {}\ \BBA {} Wohlfarth, M\BPBI N.%
\end{APACrefauthors}%
\unskip\
\newblock
\APACrefYearMonthDay{2012}{}{}.
\newblock
{\BBOQ}\APACrefatitle {{Gravitational dynamics for all tensorial spacetimes
  carrying predictive, interpretable and quantizable matter}} {{Gravitational
  dynamics for all tensorial spacetimes carrying predictive, interpretable and
  quantizable matter}}.{\BBCQ}
\newblock
\APACjournalVolNumPages{Phys. Rev. D}{85}{}{104042}.
\newblock
\begin{APACrefDOI} \doi{10.1103/PhysRevD.85.104042} \end{APACrefDOI}
\PrintBackRefs{\CurrentBib}

\bibitem [\protect \citeauthoryear {%
Giulini%
}{%
Giulini%
}{%
{\protect \APACyear {1995}}%
}]{%
Giulini1995}
\APACinsertmetastar {%
Giulini1995}%
\begin{APACrefauthors}%
Giulini, D.%
\end{APACrefauthors}%
\unskip\
\newblock
\APACrefYearMonthDay{1995}{}{}.
\newblock
{\BBOQ}\APACrefatitle {{What is the geometry of superspace?}} {{What is the
  geometry of superspace?}}{\BBCQ}
\newblock
\APACjournalVolNumPages{Phys. Rev.}{D51}{}{5630-5635}.
\newblock
\begin{APACrefDOI} \doi{10.1103/PhysRevD.51.5630} \end{APACrefDOI}
\PrintBackRefs{\CurrentBib}

\bibitem [\protect \citeauthoryear {%
Gomes%
}{%
Gomes%
}{%
{\protect \APACyear {2018}}%
}]{%
Gomes2017}
\APACinsertmetastar {%
Gomes2017}%
\begin{APACrefauthors}%
Gomes, H.%
\end{APACrefauthors}%
\unskip\
\newblock
\APACrefYearMonthDay{2018}{}{}.
\newblock
{\BBOQ}\APACrefatitle {Local gravity theories in conformal superspace} {Local
  gravity theories in conformal superspace}.{\BBCQ}
\newblock
\APACjournalVolNumPages{Annals of Physics}{}{}{}.
\newblock
\begin{APACrefURL}
  \url{http://www.sciencedirect.com/science/article/pii/S0003491618301507}
  \end{APACrefURL}
\newblock
\begin{APACrefDOI} \doi{https://doi.org/10.1016/j.aop.2018.05.014}
  \end{APACrefDOI}
\PrintBackRefs{\CurrentBib}

\bibitem [\protect \citeauthoryear {%
Gomes%
, Gryb%
\BCBL {}\ \BBA {} Koslowski%
}{%
Gomes%
\ \protect \BOthers {.}}{%
{\protect \APACyear {2011}}%
}]{%
SD_first}
\APACinsertmetastar {%
SD_first}%
\begin{APACrefauthors}%
Gomes, H.%
, Gryb, S.%
\BCBL {}\ \BBA {} Koslowski, T.%
\end{APACrefauthors}%
\unskip\
\newblock
\APACrefYearMonthDay{2011}{}{}.
\newblock
{\BBOQ}\APACrefatitle {{Einstein gravity as a 3D conformally invariant theory}}
  {{Einstein gravity as a 3D conformally invariant theory}}.{\BBCQ}
\newblock
\APACjournalVolNumPages{Class. Quant. Grav.}{28}{}{045005}.
\newblock
\begin{APACrefDOI} \doi{10.1088/0264-9381/28/4/045005} \end{APACrefDOI}
\PrintBackRefs{\CurrentBib}

\bibitem [\protect \citeauthoryear {%
Gomes%
\ \BBA {} Koslowski%
}{%
Gomes%
\ \BBA {} Koslowski%
}{%
{\protect \APACyear {2012}}%
}]{%
SD_linking}
\APACinsertmetastar {%
SD_linking}%
\begin{APACrefauthors}%
Gomes, H.%
\BCBT {}\ \BBA {} Koslowski, T.%
\end{APACrefauthors}%
\unskip\
\newblock
\APACrefYearMonthDay{2012}{}{}.
\newblock
{\BBOQ}\APACrefatitle {{The Link between General Relativity and Shape
  Dynamics}} {{The Link between General Relativity and Shape Dynamics}}.{\BBCQ}
\newblock
\APACjournalVolNumPages{Class.Quant.Grav.}{29}{}{075009}.
\newblock
\begin{APACrefDOI} \doi{10.1088/0264-9381/29/7/075009} \end{APACrefDOI}
\PrintBackRefs{\CurrentBib}

\bibitem [\protect \citeauthoryear {%
Gomes%
\ \BBA {} Shyam%
}{%
Gomes%
\ \BBA {} Shyam%
}{%
{\protect \APACyear {2016}}%
}]{%
GomesShyam}
\APACinsertmetastar {%
GomesShyam}%
\begin{APACrefauthors}%
Gomes, H.%
\BCBT {}\ \BBA {} Shyam, V.%
\end{APACrefauthors}%
\unskip\
\newblock
\APACrefYearMonthDay{2016}{}{}.
\newblock
{\BBOQ}\APACrefatitle {{Extending the rigidity of general relativity}}
  {{Extending the rigidity of general relativity}}.{\BBCQ}
\newblock
\APACjournalVolNumPages{J. Math. Phys.}{57}{11}{112503}.
\newblock
\begin{APACrefDOI} \doi{10.1063/1.4967951} \end{APACrefDOI}
\PrintBackRefs{\CurrentBib}

\bibitem [\protect \citeauthoryear {%
Gourgoulhon%
}{%
Gourgoulhon%
}{%
{\protect \APACyear {2007}}%
}]{%
3+1_book}
\APACinsertmetastar {%
3+1_book}%
\begin{APACrefauthors}%
Gourgoulhon, E.%
\end{APACrefauthors}%
\unskip\
\newblock
\APACrefYearMonthDay{2007}{}{}.
\newblock
{\BBOQ}\APACrefatitle {{3+1 formalism and bases of numerical relativity}} {{3+1
  formalism and bases of numerical relativity}}.{\BBCQ}
\newblock
\APACjournalVolNumPages{Lecture notes in Physics 846, Springer}{}{}{}.
\PrintBackRefs{\CurrentBib}

\bibitem [\protect \citeauthoryear {%
Hojman%
, Kuchar%
\BCBL {}\ \BBA {} Teitelboim%
}{%
Hojman%
\ \protect \BOthers {.}}{%
{\protect \APACyear {1976}}%
}]{%
HKT}
\APACinsertmetastar {%
HKT}%
\begin{APACrefauthors}%
Hojman, S\BPBI A.%
, Kuchar, K.%
\BCBL {}\ \BBA {} Teitelboim, C.%
\end{APACrefauthors}%
\unskip\
\newblock
\APACrefYearMonthDay{1976}{}{}.
\newblock
{\BBOQ}\APACrefatitle {{Geometrodynamics Regained}} {{Geometrodynamics
  Regained}}.{\BBCQ}
\newblock
\APACjournalVolNumPages{Annals Phys.}{96}{}{88-135}.
\newblock
\begin{APACrefDOI} \doi{10.1016/0003-4916(76)90112-3} \end{APACrefDOI}
\PrintBackRefs{\CurrentBib}

\bibitem [\protect \citeauthoryear {%
Horava%
}{%
Horava%
}{%
{\protect \APACyear {2009}}%
}]{%
Horava}
\APACinsertmetastar {%
Horava}%
\begin{APACrefauthors}%
Horava, P.%
\end{APACrefauthors}%
\unskip\
\newblock
\APACrefYearMonthDay{2009}{}{}.
\newblock
{\BBOQ}\APACrefatitle {{Quantum Gravity at a Lifshitz Point}} {{Quantum Gravity
  at a Lifshitz Point}}.{\BBCQ}
\newblock
\APACjournalVolNumPages{Phys.Rev.}{D79}{}{084008}.
\newblock
\begin{APACrefDOI} \doi{10.1103/PhysRevD.79.084008} \end{APACrefDOI}
\PrintBackRefs{\CurrentBib}

\bibitem [\protect \citeauthoryear {%
Huggett%
\ \BBA {} W\"uthrich%
}{%
Huggett%
\ \BBA {} W\"uthrich%
}{%
{\protect \APACyear {2021}}%
}]{%
HuggettWuttrich_book}
\APACinsertmetastar {%
HuggettWuttrich_book}%
\begin{APACrefauthors}%
Huggett, N.%
\BCBT {}\ \BBA {} W\"uthrich, C.%
\end{APACrefauthors}%
\unskip\
\newblock
\APACrefYear{2021}.
\newblock
\APACrefbtitle {{Out of Nowhere: The Emergence of Spacetime in Quantum Theories
  of Gravity}} {{Out of Nowhere: The Emergence of Spacetime in Quantum Theories
  of Gravity}}.
\newblock
\APACaddressPublisher{}{Oxford University Press.}
\PrintBackRefs{\CurrentBib}

\bibitem [\protect \citeauthoryear {%
Isenberg%
, Murchadha%
\BCBL {}\ \BBA {} J.~W.~York%
}{%
Isenberg%
\ \protect \BOthers {.}}{%
{\protect \APACyear {1976}}%
}]{%
York3}
\APACinsertmetastar {%
York3}%
\begin{APACrefauthors}%
Isenberg, J\BPBI A.%
, Murchadha, N.%
\BCBL {}\ \BBA {} J.~W.~York, J.%
\end{APACrefauthors}%
\unskip\
\newblock
\APACrefYearMonthDay{1976}{}{}.
\newblock
{\BBOQ}\APACrefatitle {{Initial Value Problem of General Relativity. 3. Coupled
  Fields and the Scalar-Tensor Theory.}} {{Initial Value Problem of General
  Relativity. 3. Coupled Fields and the Scalar-Tensor Theory.}}{\BBCQ}
\newblock
\APACjournalVolNumPages{Phys. Rev. D , 13, 1532-1537.}{}{}{}.
\PrintBackRefs{\CurrentBib}

\bibitem [\protect \citeauthoryear {%
James%
}{%
James%
}{%
{\protect \APACyear {2020}}%
}]{%
James2020}
\APACinsertmetastar {%
James2020}%
\begin{APACrefauthors}%
James, L.%
\end{APACrefauthors}%
\unskip\
\newblock
\APACrefYearMonthDay{2020}{}{}.
\newblock
{\BBOQ}\APACrefatitle {A new perspective on time and physical laws} {A new
  perspective on time and physical laws}.{\BBCQ}
\newblock
\APACjournalVolNumPages{{\em Forthcoming in {\em British Journal of Philosophy
  of Science}: http://philsci-archive.pitt.edu/17107/}}{}{}{}.
\PrintBackRefs{\CurrentBib}

\bibitem [\protect \citeauthoryear {%
Kiefer%
}{%
Kiefer%
}{%
{\protect \APACyear {2012}}%
}]{%
Kiefer_book}
\APACinsertmetastar {%
Kiefer_book}%
\begin{APACrefauthors}%
Kiefer, K.%
\end{APACrefauthors}%
\unskip\
\newblock
\APACrefYear{2012}.
\newblock
\APACrefbtitle {{Quantum Gravity}} {{Quantum Gravity}}.
\newblock
\APACaddressPublisher{}{Oxford University Press}.
\PrintBackRefs{\CurrentBib}

\bibitem [\protect \citeauthoryear {%
Lam%
\ \BBA {} W\"uthrich%
}{%
Lam%
\ \BBA {} W\"uthrich%
}{%
{\protect \APACyear {2020}}%
}]{%
LamWuttrich_func}
\APACinsertmetastar {%
LamWuttrich_func}%
\begin{APACrefauthors}%
Lam, V.%
\BCBT {}\ \BBA {} W\"uthrich, C.%
\end{APACrefauthors}%
\unskip\
\newblock
\APACrefYearMonthDay{2020}{}{}.
\newblock
{\BBOQ}\APACrefatitle {Spacetime functionalism from a realist perspective}
  {Spacetime functionalism from a realist perspective}.{\BBCQ}
\newblock
\APACjournalVolNumPages{Synthese}{}{}{}.
\PrintBackRefs{\CurrentBib}

\bibitem [\protect \citeauthoryear {%
Lewis%
}{%
Lewis%
}{%
{\protect \APACyear {1970}}%
}]{%
Lewis_defT}
\APACinsertmetastar {%
Lewis_defT}%
\begin{APACrefauthors}%
Lewis, D.%
\end{APACrefauthors}%
\unskip\
\newblock
\APACrefYearMonthDay{1970}{}{}.
\newblock
{\BBOQ}\APACrefatitle {{How to Define Theoretical Terms}} {{How to Define
  Theoretical Terms}}.{\BBCQ}
\newblock
\APACjournalVolNumPages{Journal of Philosophy, {\bf 67}, {\em
  pp.~427-446.}}{}{}{}.
\PrintBackRefs{\CurrentBib}

\bibitem [\protect \citeauthoryear {%
Lewis%
}{%
Lewis%
}{%
{\protect \APACyear {1972}}%
}]{%
Lewis_func}
\APACinsertmetastar {%
Lewis_func}%
\begin{APACrefauthors}%
Lewis, D.%
\end{APACrefauthors}%
\unskip\
\newblock
\APACrefYearMonthDay{1972}{}{}.
\newblock
{\BBOQ}\APACrefatitle {Psychophysical and theoretical identifications}
  {Psychophysical and theoretical identifications}.{\BBCQ}
\newblock
\APACjournalVolNumPages{Australasian Journal of Philosophy, {vol. {\bf 50},
  p.249-258}}{}{}{}.
\PrintBackRefs{\CurrentBib}

\bibitem [\protect \citeauthoryear {%
Lovelock%
}{%
Lovelock%
}{%
{\protect \APACyear {1971}}%
}]{%
Lovelock}
\APACinsertmetastar {%
Lovelock}%
\begin{APACrefauthors}%
Lovelock, D.%
\end{APACrefauthors}%
\unskip\
\newblock
\APACrefYearMonthDay{1971}{}{}.
\newblock
{\BBOQ}\APACrefatitle {{The Einstein Tensor and Its Generalizations}} {{The
  Einstein Tensor and Its Generalizations}}.{\BBCQ}
\newblock
\APACjournalVolNumPages{Journal of Mathematical Physics}{12}{3}{498-501}.
\newblock
\begin{APACrefURL} \url{https://doi.org/10.1063/1.1665613} \end{APACrefURL}
\newblock
\begin{APACrefDOI} \doi{10.1063/1.1665613} \end{APACrefDOI}
\PrintBackRefs{\CurrentBib}

\bibitem [\protect \citeauthoryear {%
Marsden%
\ \BBA {} Tipler%
}{%
Marsden%
\ \BBA {} Tipler%
}{%
{\protect \APACyear {1980}}%
}]{%
Tipler_CMC}
\APACinsertmetastar {%
Tipler_CMC}%
\begin{APACrefauthors}%
Marsden, J\BPBI E.%
\BCBT {}\ \BBA {} Tipler, F\BPBI J.%
\end{APACrefauthors}%
\unskip\
\newblock
\APACrefYearMonthDay{1980}{}{}.
\newblock
{\BBOQ}\APACrefatitle {{Maximal hypersurfaces and foliations of constant mean
  extrinsic curvature in General Relativity}} {{Maximal hypersurfaces and
  foliations of constant mean extrinsic curvature in General
  Relativity}}.{\BBCQ}
\newblock
\APACjournalVolNumPages{{Physics Reports}}{\bf 66}{}{109}.
\PrintBackRefs{\CurrentBib}

\bibitem [\protect \citeauthoryear {%
McTaggart%
}{%
McTaggart%
}{%
{\protect \APACyear {1908}}%
}]{%
Mctaggart}
\APACinsertmetastar {%
Mctaggart}%
\begin{APACrefauthors}%
McTaggart, J\BPBI E.%
\end{APACrefauthors}%
\unskip\
\newblock
\APACrefYearMonthDay{1908}{}{}.
\newblock
{\BBOQ}\APACrefatitle {The unreality of time} {The unreality of time}.{\BBCQ}
\newblock
\APACjournalVolNumPages{{\em {\em Mind}, {\bf 17}, pp.~457-474.}}{}{}{}.
\PrintBackRefs{\CurrentBib}

\bibitem [\protect \citeauthoryear {%
Menon%
}{%
Menon%
}{%
{\protect \APACyear {2021}}%
}]{%
Menon_lor}
\APACinsertmetastar {%
Menon_lor}%
\begin{APACrefauthors}%
Menon, T.%
\end{APACrefauthors}%
\unskip\
\newblock
\APACrefYearMonthDay{2021}{}{}.
\newblock
{\BBOQ}\APACrefatitle {{Why is the metric of general relativity Lorentzian?}}
  {{Why is the metric of general relativity Lorentzian?}}{\BBCQ}
\newblock
\APACjournalVolNumPages{In preparation}{}{}{}.
\PrintBackRefs{\CurrentBib}

\bibitem [\protect \citeauthoryear {%
Mercati%
}{%
Mercati%
}{%
{\protect \APACyear {2017}}%
}]{%
Flavio_tutorial}
\APACinsertmetastar {%
Flavio_tutorial}%
\begin{APACrefauthors}%
Mercati, F.%
\end{APACrefauthors}%
\unskip\
\newblock
\APACrefYear{2017}.
\newblock
\APACrefbtitle {{Shape Dynamics: Relativity and Relationalism}} {{Shape
  Dynamics: Relativity and Relationalism}}.
\newblock
\APACaddressPublisher{}{Oxford University Press}.
\PrintBackRefs{\CurrentBib}

\bibitem [\protect \citeauthoryear {%
Nagel%
}{%
Nagel%
}{%
{\protect \APACyear {1961}}%
}]{%
Nagel_book}
\APACinsertmetastar {%
Nagel_book}%
\begin{APACrefauthors}%
Nagel, E.%
\end{APACrefauthors}%
\unskip\
\newblock
\APACrefYear{1961}.
\newblock
\APACrefbtitle {{The Structure of Science: Problems in the logic of scientific
  explanation}} {{The Structure of Science: Problems in the logic of scientific
  explanation}}.
\newblock
\APACaddressPublisher{}{Harcourt}.
\PrintBackRefs{\CurrentBib}

\bibitem [\protect \citeauthoryear {%
Nagel%
}{%
Nagel%
}{%
{\protect \APACyear {2008}}%
}]{%
Nagel_1979}
\APACinsertmetastar {%
Nagel_1979}%
\begin{APACrefauthors}%
Nagel, E.%
\end{APACrefauthors}%
\unskip\
\newblock
\APACrefYearMonthDay{2008}{}{}.
\newblock
{\BBOQ}\APACrefatitle {Issues in the logic of reductive explanations (1979)}
  {Issues in the logic of reductive explanations (1979)}.{\BBCQ}
\newblock
\BIn{} \APACrefbtitle {{Emergence: Contemporary Readings, {\em edited by M.
  Bedau and P. Humphreys}}.} {{Emergence: Contemporary Readings, {\em edited by
  M. Bedau and P. Humphreys}}.}
\newblock
\APACaddressPublisher{}{MIT press}.
\PrintBackRefs{\CurrentBib}

\bibitem [\protect \citeauthoryear {%
Oliver%
\ \BBA {} Smiley%
}{%
Oliver%
\ \BBA {} Smiley%
}{%
{\protect \APACyear {2016}}%
}]{%
OliverSmiley}
\APACinsertmetastar {%
OliverSmiley}%
\begin{APACrefauthors}%
Oliver, A.%
\BCBT {}\ \BBA {} Smiley, T.%
\end{APACrefauthors}%
\unskip\
\newblock
\APACrefYear{2016}.
\newblock
\APACrefbtitle {{Plural Logic}} {{Plural Logic}}.
\newblock
\APACaddressPublisher{}{Oxford University Press.}
\PrintBackRefs{\CurrentBib}

\bibitem [\protect \citeauthoryear {%
Schaffner%
}{%
Schaffner%
}{%
{\protect \APACyear {2012}}%
}]{%
SchaffnerNagel2012}
\APACinsertmetastar {%
SchaffnerNagel2012}%
\begin{APACrefauthors}%
Schaffner, K.%
\end{APACrefauthors}%
\unskip\
\newblock
\APACrefYearMonthDay{2012}{}{}.
\newblock
{\BBOQ}\APACrefatitle {{Ernest Nagel and reduction}} {{Ernest Nagel and
  reduction}}.{\BBCQ}
\newblock
\APACjournalVolNumPages{Journal of Philosophy, {\em {\bf 109}, 534-565}}{}{}{}.
\PrintBackRefs{\CurrentBib}

\bibitem [\protect \citeauthoryear {%
Schuller%
}{%
Schuller%
}{%
{\protect \APACyear {2011}}%
}]{%
Schuller}
\APACinsertmetastar {%
Schuller}%
\begin{APACrefauthors}%
Schuller, F\BPBI P.%
\end{APACrefauthors}%
\unskip\
\newblock
\APACrefYearMonthDay{2011}{}{}.
\newblock
{\BBOQ}\APACrefatitle {{All spacetimes beyond Einstein (Obergurgl Lectures)}}
  {{All spacetimes beyond Einstein (Obergurgl Lectures)}}.{\BBCQ}
\newblock
\APACjournalVolNumPages{arXiv:1111.4824}{}{}{}.
\PrintBackRefs{\CurrentBib}

\bibitem [\protect \citeauthoryear {%
Stelle%
}{%
Stelle%
}{%
{\protect \APACyear {1977}}%
}]{%
Stelle_grav}
\APACinsertmetastar {%
Stelle_grav}%
\begin{APACrefauthors}%
Stelle, K\BPBI S.%
\end{APACrefauthors}%
\unskip\
\newblock
\APACrefYearMonthDay{1977}{Aug}{}.
\newblock
{\BBOQ}\APACrefatitle {Renormalization of higher-derivative quantum gravity}
  {Renormalization of higher-derivative quantum gravity}.{\BBCQ}
\newblock
\APACjournalVolNumPages{Phys. Rev. D}{16}{}{953--969}.
\newblock
\begin{APACrefURL} \url{https://link.aps.org/doi/10.1103/PhysRevD.16.953}
  \end{APACrefURL}
\newblock
\begin{APACrefDOI} \doi{10.1103/PhysRevD.16.953} \end{APACrefDOI}
\PrintBackRefs{\CurrentBib}

\bibitem [\protect \citeauthoryear {%
Teitelboim%
}{%
Teitelboim%
}{%
{\protect \APACyear {1973}}%
}]{%
Teitelboim1973}
\APACinsertmetastar {%
Teitelboim1973}%
\begin{APACrefauthors}%
Teitelboim, C.%
\end{APACrefauthors}%
\unskip\
\newblock
\APACrefYearMonthDay{1973}{}{}.
\newblock
{\BBOQ}\APACrefatitle {How commutators of constraints reflect the spacetime
  structure} {How commutators of constraints reflect the spacetime
  structure}.{\BBCQ}
\newblock
\APACjournalVolNumPages{Annals of Physics}{79}{2}{542 - 557}.
\newblock
\begin{APACrefURL}
  \url{http://www.sciencedirect.com/science/article/pii/0003491673900961}
  \end{APACrefURL}
\newblock
\begin{APACrefDOI} \doi{https://doi.org/10.1016/0003-4916(73)90096-1}
  \end{APACrefDOI}
\PrintBackRefs{\CurrentBib}

\bibitem [\protect \citeauthoryear {%
Woodard%
}{%
Woodard%
}{%
{\protect \APACyear {2015}}%
}]{%
Woodard_scholarpedia}
\APACinsertmetastar {%
Woodard_scholarpedia}%
\begin{APACrefauthors}%
Woodard, R\BPBI P.%
\end{APACrefauthors}%
\unskip\
\newblock
\APACrefYearMonthDay{2015}{}{}.
\newblock
{\BBOQ}\APACrefatitle {{Ostrogradsky's theorem on Hamiltonian instability}}
  {{Ostrogradsky's theorem on Hamiltonian instability}}.{\BBCQ}
\newblock
\APACjournalVolNumPages{Scholarpedia, 10(8):32243.}{}{}{}.
\PrintBackRefs{\CurrentBib}

\bibitem [\protect \citeauthoryear {%
York%
}{%
York%
}{%
{\protect \APACyear {1971}}%
}]{%
York}
\APACinsertmetastar {%
York}%
\begin{APACrefauthors}%
York, J\BPBI W.%
\end{APACrefauthors}%
\unskip\
\newblock
\APACrefYearMonthDay{1971}{}{}.
\newblock
{\BBOQ}\APACrefatitle {Gravitational degrees of freedom and the initial-value
  problem} {Gravitational degrees of freedom and the initial-value
  problem}.{\BBCQ}
\newblock
\APACjournalVolNumPages{Phys. Rev. Lett.}{26}{}{1656--1658}.
\newblock
\begin{APACrefDOI} \doi{10.1103/PhysRevLett.26.1656} \end{APACrefDOI}
\PrintBackRefs{\CurrentBib}

\bibitem [\protect \citeauthoryear {%
Zeh%
}{%
Zeh%
}{%
{\protect \APACyear {1992}}%
}]{%
Zeh_time}
\APACinsertmetastar {%
Zeh_time}%
\begin{APACrefauthors}%
Zeh, H\BHBI D.%
\end{APACrefauthors}%
\unskip\
\newblock
\APACrefYear{1992}.
\newblock
\APACrefbtitle {{The Physical Basis of the Direction of Time}} {{The Physical
  Basis of the Direction of Time}}.
\newblock
\APACaddressPublisher{}{Springer}.
\PrintBackRefs{\CurrentBib}

\end{thebibliography}

\end{document}